\documentclass[10pt,journal,compsoc]{IEEEtran}

\ifCLASSOPTIONcompsoc
  \usepackage[nocompress]{cite}
\else
  \usepackage{cite}
\fi

\ifCLASSINFOpdf
   \usepackage[pdftex]{graphicx}

\else

   \usepackage[dvips]{graphicx}

\fi

\usepackage{textcomp}
\usepackage{multirow} 
\usepackage{verbatim}
\usepackage{threeparttable}
\usepackage{makecell}
\usepackage{graphicx}
\usepackage{textcomp}
\usepackage{bbding}
\usepackage{pifont}
\usepackage{array} 
\usepackage{longtable}
\usepackage{amsthm,amsmath,amssymb}
\usepackage{booktabs}
\usepackage{float}
\usepackage[small]{caption}
\usepackage[hidelinks]{hyperref}
\usepackage{amsmath}

\usepackage{algorithmic}
\usepackage{algorithm}
\makeatletter
\newcommand{\removelatexerror}{\let\@latex@error\@gobble}
\makeatother

  \usepackage[caption=false,font=footnotesize,labelfont=sf,textfont=sf]{subfig}
  \usepackage[caption=false,font=footnotesize]{subfig}

\begin{document}

\title{DIP-Watermark: A Double Identity Protection Method Based on Robust Adversarial Watermark}

\author{ Yunming Zhang, 
    Dengpan Ye,~\IEEEmembership{Member~IEEE,}
  Caiyun Xie, Sipeng Shen, Ziyi Liu, Jiacheng Deng, Long Tang
\IEEEcompsocitemizethanks{\IEEEcompsocthanksitem 
 Yunming Zhang, Dengpan Ye, Caiyun Xie, Sipeng Shen, Ziyi Liu, Jiacheng Deng and Long Tang are with the Key Laboratory of Aerospace Information Security and Trusted Computing, Ministry of Education, School of Cyber Science and Engineering, Wuhan University. E-mail: {\{Yunming Zhang, yedp, caiyunxie, ziyi$\_$liu,  dengjiacheng, l$\_$tang\}}@whu.edu.cn, 1292365221shen@gmail.com.\\
(Corresponding author: Dengpan Ye)}}

\markboth{Journal of \LaTeX\ Class Files}%
{Shell \MakeLowercase{\textit{et al.}}: Rethinking Adversarial Robustness through a Perspective of Attention}

\IEEEtitleabstractindextext{%
\begin{abstract}
The wide deployment of Face Recognition (FR) systems poses privacy risks. One countermeasure is adversarial attack, deceiving unauthorized malicious FR, but it also disrupts regular identity verification of trusted authorizers, exacerbating the potential threat of identity impersonation. To address this, we propose the first double identity protection scheme based on traceable adversarial watermarking, termed DIP-Watermark. DIP-Watermark employs a one-time watermark embedding to deceive unauthorized FR models and allows authorizers to perform identity verification by extracting the watermark. Specifically, we propose an information-guided adversarial attack against FR models. The encoder embeds an identity-specific watermark into the deep feature space of the carrier, guiding recognizable features of the image to deviate from the source identity. We further adopt a collaborative meta-optimization strategy compatible with sub-tasks, which regularizes the joint optimization direction of the encoder and decoder. This strategy enhances the representation of universal carrier features, mitigating multi-objective optimization conflicts in watermarking. Experiments confirm that DIP-Watermark achieves significant attack success rates and traceability accuracy on state-of-the-art FR models, exhibiting remarkable robustness that outperforms the existing privacy protection methods using adversarial attacks and deep watermarking, or simple combinations of the two. Our work potentially opens up new insights into proactive protection for FR privacy.
\end{abstract}

}

\maketitle

\IEEEdisplaynontitleabstractindextext

\IEEEpeerreviewmaketitle

\IEEEraisesectionheading{\section{Introduction}\label{sec:introduction}}

\IEEEPARstart{B}{enefiting} from the rapid development of deep learning and cloud storage technologies, Face Recognition (FR) systems have experienced significant development and widely deployed in the fields of criminal investigation, security monitoring, and online supervision, facilitating genuine identity verification and management based on facial images. Unfortunately, the rise of FR technology also poses a threat to user privacy~\cite{liu2023diffprotect,hu2022protecting,NEURIPS2023_29e4b51d,9633176,9442369}. Unauthorized Adversaries (UAs) can use open-source FR recognition systems to match faces obtained from online platforms, thereby stealing users' real identities and disclosing private information.

\begin{figure}
	\centering
	\includegraphics[width=85mm]{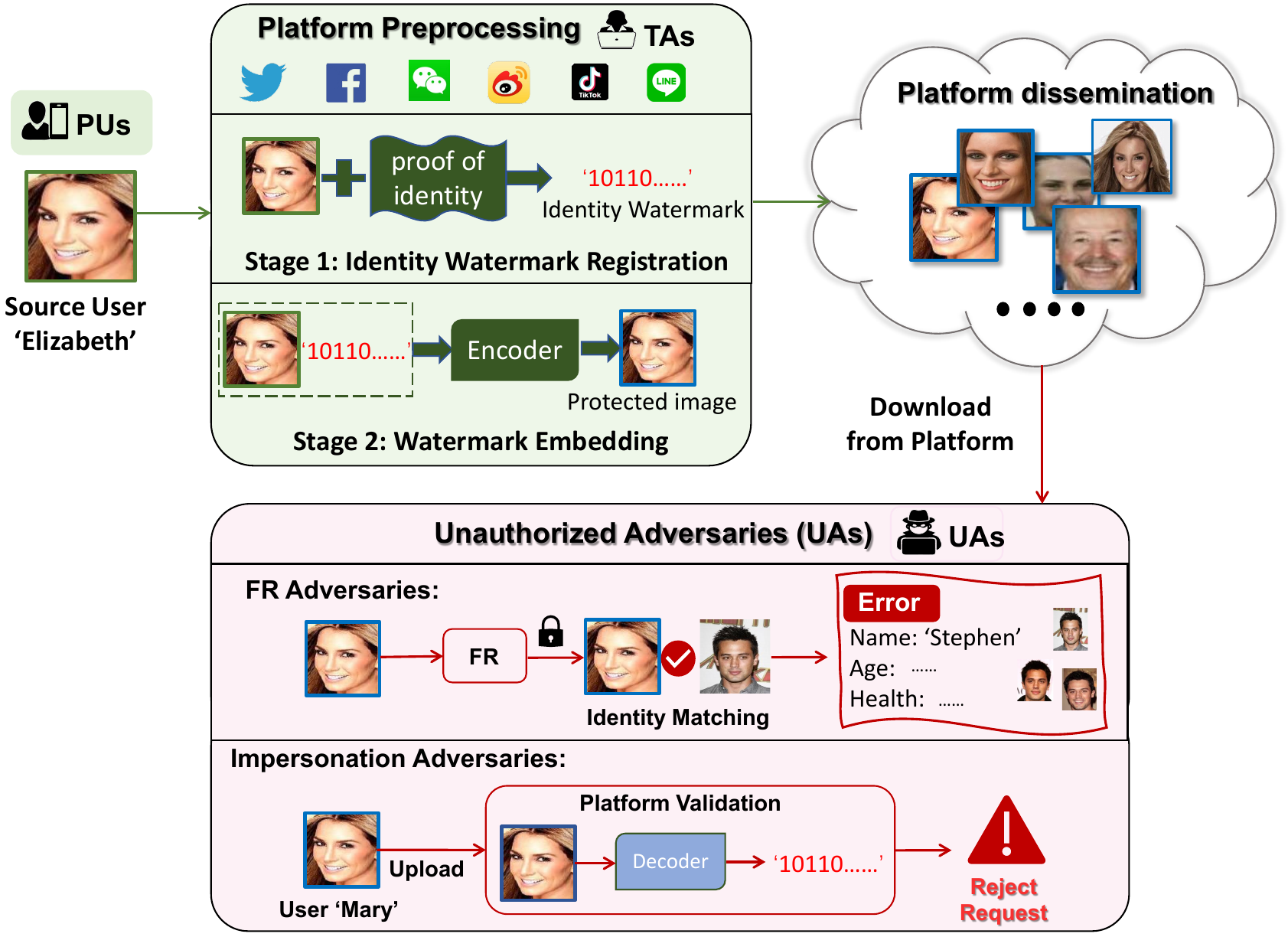}
        \caption{Illustration of DIP-Watermark protection scenario. Before users upload personal images, the platform verifies their real identity and registers an identity-specific watermark. After embedding the watermark, the platform uploads the watermarked image for sharing, while retaining the watermark decoder, which is authorized to regulatory bodies. When malicious FR adversaries attempt FR on the protected user image downloaded from the platform, the adversariality of the image disrupts the FR model, leading to incorrect matching results. In cases where impersonators attempt to use user images for identity theft while using the platform's services, the platform employs the decoder for identity verification and rejects the request. Meanwhile, TAs can use the authorized decoder for image identity verification when necessary.}	
 \label{fig1}
\end{figure}

Several identity privacy protection schemes based on adversarial attacks have been proposed to address the unauthorized malicious Facial Recognition (FR) issue~\cite{hu2022protecting, komkov2021advhat,jia2022adv}. These methods generate adversarial facial images by adding imperceptible adversarial perturbations~\cite{Li_2023_CVPR,Dong2019EfficientDB} or visible adversarial patches to images~\cite{yang2021attacks,wei2022simultaneously}, which disrupt the decision-making of FR models and lead them to match incorrect facial images, thus protecting real identity information. However, these methods exhibit certain drawbacks, including poor black-box transferability, lack of robustness, and significant semantic modifications to protected images~\cite{jia2022adv}. Most Importantly, these methods also raise additional privacy concerns. Malicious users may exploit adversarial images for identity impersonation, attempting to use platform services under an assumed identity. Since adversarial examples have lost recognizable features, even Trusted Authorizers (TAs) cannot conduct legitimate and compliant identity verification through FR, affecting normal security supervision. Therefore, there is an urgent need for a comprehensive privacy protection technique that, while countering unauthorized malicious FR, ensures compliant identity verification by TAs, thus avoiding the potential threat of identity impersonation.


To address the limitations of current identity privacy protection schemes, this paper introduces an innovative problem: Can we transform adversarial examples into verifiable patterns for identity verification of TAs while deceiving FR models of UAs? Deep watermarking methods offer a new perspective for verifiable adversarial examples~\cite{zhu2018hidden}, which typically protect the copyright of digital media by invisibly embedding traceable watermarks. The watermark encoder learns the embedding of watermarks in the deep feature space of images to enhance robustness, enabling deployment in complex social network scenarios. However, traditional deep watermark algorithms fail to disrupt malicious face recognition, thus unable to fundamentally address privacy leakage issues.

In this paper, we propose the first Dual Identity Protection scheme based on robust and traceable adversarial watermarking, termed \ ``DIP-Watermark." DIP-Watermark serves as an adversarial deep watermark that protects user identity privacy from the perspectives of disrupting unauthorized malicious FR models and real identity traceability. As illustrated in Fig. 1, when users upload personal facial images to online platforms, the platform embed identity-specific adversarial watermarks in user images. The embedded watermark does not affect the visual appearance or social usability of the image. Malicious adversaries may collect user images embedded with adversarial watermarks without permission and attempt to disclose users' real identity information through FR systems. However, the adversarial watermark disrupts the FR model so that adversaries get wrong matching information, thus safeguarding user privacy. Meanwhile, in processes such as copyright claims, online supervision, or criminal investigations, authorized platforms can extract the watermark using the watermark decoder to verify the real identity information of the facial image, thereby mitigating the potential threat of identity impersonation. The comparison between DIP-Watermark and existing works is illustrated in Table~\ref{Func com}.  \ ``Original" denotes the original scenario, indicating that the generated adversarial samples or watermark images have not undergone any image post-processing. \ ``Transferability" indicates the extent to which the method exhibits transferability. \ ``Robust" represents the robust scenario, indicating that the generated images have undergone image post-processing. Checkmarks and crosses indicate whether the method has the performance in the corresponding scenarios.

DIP-Watermark focuses on the following three challenges to achieve comprehensive privacy protection: \textit{\textbf{1) Traceable and adversarial dual-effectiveness}}: We introduce an information-guided adversarial attack strategy against FR models. This strategy embeds an identity-specific watermark into the source image using the encoder, guiding the recognizable features of the source image to map onto the target feature domain, deceiving FR models. \textit{\textbf{2) Transferability to black-box FR models}}: There is an optimization conflict between the adversariality and invisibility of watermarks. Adversariality enhancements under single-model or basic ensemble overly penalize image quality and lack transferability to black-box FR models. Therefore, we adopt a meta-learning strategy in the multi-task end-to-end optimization, calibrating training errors across distinct tasks, thereby enhancing the encoding of universal carrier features. \textit{\textbf{3) Model robustness against common image processing}}: We add a simulated noise pool and design regularization under temporary encoder parameters to adjust the optimization direction of the decoder. This enhances the robustness of the decoder and facilitates the synergistic optimization between the encoder and decoder. Furthermore, multi-task training contributes to the overall robustness of the model.

Our contributions are summarized as follows:
\begin{itemize}
\item  We propose DIP-Watermark, the first dual-effect identity protection scheme based on traceable adversarial watermark. It can simultaneously disrupt unauthorized FR models and enable TAs to perform normal image identity verification tasks.



\item  We innovatively present an information-guided adversarial attack method against FR models, which maps recognizable features of the image to the target domain through identity-specific watermarks, influencing the facial feature reconstruction of FR models.

\item We adopt the collaborative meta-optimization strategy compatible with diverse tasks to enhance the universal feature encoding of the carrier, mitigating optimization conflicts between adversariality and invisibility.



\item 
We conduct extensive experiments on two large facial datasets against six SOTA offline FR models and two online commercial FR systems, validating the significant transferability, traceability and robustness of  DIP-Watermark.
\end{itemize}
\section{Related Works}
\subsection{Adversarial Attacks against Face Recognition}

Studies have demonstrated the vulnerability of FR models to adversarial examples. White-box attacks~\cite{sharif2019general} and query-based black-box attacks~\cite{wei2022simultaneously} heavily depend on model accessibility, rendering them impractical for real-world applications. Therefore, more privacy protection methods are focusing on black-box attacks for FR models. Sharif et al.~\cite{10.1145/2976749.2978392} propose the first patch-based attacks, which restricts the perturbation to eyeglasses and physically realizes attacks on FR models. Zhu et al.~\cite{8803269} synthesize eye makeup patches to attack FR models, and Yin et al.~\cite{yin2021adv} further improve the realism of eye makeup and enhanced black-box transferability. Komkov et al.~\cite{komkov2021advhat} propose AdvHat, a real-world adversarial attack using sticker on the hat. To eliminate vulnerabilities arising from low-level pixel modifications and enhance transferability to black-box FR models, some studies design perturbation-based attack methods with PGD~\cite{madry2017towards} or FGSM~\cite{goodfellow2014explaining} as underlying optimization algorithms. Zhong et al.~\cite{zhong2020towards} increase the diversity of surrogate models and obtain ensemble-like effects, significantly improving the transferability. Li et al.~\cite{Li_2023_CVPR} enhances transferability by selecting a attribute recognition task related to FR. However, the perturbation-based methods directly overlay gradients in the pixel domain of the carrier, lacking robustness. In essence, within the context of identity privacy protection, adversarial attacks irreversibly disrupt recognizable features essential for FR, thereby hindering normal identity verification by authorizers and impacting the compliant security oversight, potentially exacerbating the threat of identity impersonation.

\begin{table}[t]

\centering

 \resizebox{\hsize}{!}{
\begin{tabular}{c|c|ccc|ccccc}

			\Xhline{1px}
			\multirow{2}{*}{\textbf{Method}} &\multirow{2}{*}{\textbf{Type}} & \multicolumn{3}{c}{\textbf{Adversariality}} & \multicolumn{2}{c}{\textbf{Traceability}}\\ \cline{3-5} \cline{6-7}
   
                &   &Original & Transferability & Robust & Original & Robust\\
                 
			\hline
			Adv-Makeup~\cite{yin2021adv} & Patch& \Checkmark&  \Checkmark & \ding{56} &\ding{56} &\ding{56}\\
                AdvHat~\cite{komkov2021advhat} & Patch& \Checkmark& \ding{56} & \ding{56} &\ding{56} &\ding{56}\\
                
              PGD~\cite{madry2017towards} & Perturbation& \Checkmark & \Checkmark& \ding{56} &\ding{56} &\ding{56}\\
              FGSM~\cite{goodfellow2014explaining} & Perturbation& \Checkmark & \Checkmark & \ding{56} &\ding{56} &\ding{56}\\
             Hidden~\cite{zhu2018hidden}& Watermark & \ding{56} & \ding{56} & \ding{56} &\Checkmark &\Checkmark \\
		FakeTagger~\cite{wang2021faketagger} & Watermark& \ding{56} & \ding{56} & \ding{56} &\Checkmark &\Checkmark\\

			  \textbf{DIP-Watermark (ours)} & Watermark & \Checkmark & \Checkmark &\Checkmark &\Checkmark &\Checkmark \\
			\Xhline{1px}
\end{tabular}}
	\caption{Comparisons of relevant research on privacy protection.}
\label{Func com}
\end{table}

\subsection{Deep Watermarking}

Deep watermarking is extensively employed for copyright tracking in digital media~\cite{zhu2018hidden, liu2019novel, 9999475}. Zhu et al.~\cite{zhu2018hidden} proposed the first end-to-end DNN-based watermarking algorithm, which accomplishes watermark embedding and extraction through an autoencoder structure and enhances watermark image quality through a discriminator. Due to complex network processing, image propagation significantly alters the data distribution of watermark information, impacting its recovery. Consequently, to enhance algorithm robustness, some deep watermarking approaches are focusing on simulated noise layer design. Liu et al.~\cite{liu2019novel} proposed a two-stage separable deep watermarking framework,  addressing the limitations of traditional end-to-end training for nondifferentiable noise. Jia et al.~\cite{jia2021mbrs} proposed a mini-batch real and simulated JPEG compression method, randomly changing the noise layer in mini batches to help the model search for optimal solutions in different directions. With increasing concerns about copyright and privacy protection in facial images, recent studies have proposed utilizing deep watermarking methods to enhance the security of facial images. Wang et al.~\cite{wang2021faketagger} utilize the deep watermarking method to encode watermarks into carrier face images, enabling effective tracing of deeply manipulated facial images. Yu et al.~\cite{Yu_2021_ICCV} introduce Artificial Fingerprints, which tracks facial images by transferring fingerprints from training data to the deep forgery model. The watermark encoder deeply embeds watermark information into the robust features of the carrier, and through end-to-end training, it learns the globally optimal solution of the encoder and decoder. This inspires us to design a dual-effect privacy protection scheme for complex social network scenarios on this basis. By defining an information-guided adversarial attack, it can avoid the irrecoverable issue of adversarial perturbations, enabling TAs to decode the watermark for identity verification. 

\section{Threat Model}
We define a threat model in terms of the potential processing of facial images in propagation, considering three entities with different knowledge and capabilities: 

\subsection{Personal Users (PUs)}
PUs possess a set of personal images $x_{s} \in \mathbb{R} ^{i\times h\times w }$ ($\mathbb{R}$ is a real field of dimension $i\times h\times w $ ) and intend to anonymously share images on public network platforms. PUs are required to register with their real identity before using platform services and have the right to request platforms to handle privacy protection of uploaded images. This is to prevent unauthorized adversaries from matching and retrieving users' real identity information through FR systems, thereby avoiding privacy breaches. Simultaneously, PUs registering on the platform implicitly authorize trusted entities, including judicial and regulatory bodies, for identity management. This enables compliant identity verification during criminal investigations or image identity copyright disputes.


\subsection{Unauthorized Adversaries (UAs)}

We define two types of UAs based on potential malicious scenarios encountered during the dissemination of facial images.
\begin{itemize}

  \item  \textit{FR adversaries: }These adversaries possess open-source FR models $\mathcal{F} (\cdot)$ and acquire facial images of users from online platforms without authorization. They attempt to retrieve the real identity of the image by face image matching through the FR model and obtain user's private information. 
  \item  \textit{Impersonation adversaries:} Malicious users may exploit other users' protected facial images for identity impersonation, either by registering on platforms with false identities or by uploading others' images to disseminate misinformation through identity impersonation. This may lead to adverse effects such as financial loss and reputational damage for the original user.
  

\end{itemize}

\subsection{Trusted Authorities (TAs)}

As trusted authorizers for user identity information, public network platforms require authentic information management of facial images. When users need to share facial images anonymously on the platform, privacy processing is required to protect the real identity information of users in the image from unauthorized adversaries. The platform can apply the DIP-Watermark method to watermark the facial images uploaded by users. This requires the platform to pre-train a pair of watermark encoder $\mathcal{E}(\cdot)$ and decoder $\mathcal{D}(\cdot)$ in the background. It is noteworthy that the platform only requires training the DIP-Watermark model once, enabling it to be utilized for embedding and extracting watermarks from images of diverse users with distinct identities. This significantly economizes computational resources of the platform. The platform can use open-source FR models during the training process, but it does not know which FR model the adversary is using; that is, the adversary's attack process is completely black-box to the platform. 

The user is first authenticated before using the platform services, and registers a unique identity-specific watermark ${{w_{id}}\in [0,1]}^{1\times L}$ ($L$ represents the length of the watermark), which is formulated by the platform. The platform then uses the watermark encoder to embed the watermark invisibly into the user image as follows:

\begin{eqnarray}\label{en}
\hat{x} _{s}=\mathcal{E}(x_{s},w_{id}).
\end{eqnarray}

Due to the invisible embedding of the watermark into the deep feature map of the image, this encoding process does not affect the image's social usability. Simultaneously, this privacy management process does not interfere with the platform's normal identity verification procedure. When malicious users attempt to deceive the platform's FR model using adversarial images, and impersonate others to use platform services, the platform can utilize the watermark decoder to extract watermark information $\hat{w} _{id}$ from the image for identity verification, thereby rejecting usage requests from malicious users with mismatched identities. The watermark decoding process is as follows:

\begin{eqnarray}\label{de}
\hat{w} _{id}=\mathcal{D}(\hat{x}_{s}).
\end{eqnarray}

Given the threat scenario, the practical application of DIP-Watermark necessitates optimization training in three aspects: \textit{(1) Invisibility optimization}: Unlike adversarial patches and face anonymization, DIP aims to ensure that watermark embedding does not significantly affect the visual quality of the carrier, maintaining the social usability of the facial image, making it more suitable for practical application requirements; \textit{(2) Adversarial optimization}: Since in practice, the network platform does not know the FR model used by the FR adversary, the watermarked image must possess adversarial transferability to disrupt unknown FR models; \textit{(3) Traceability optimization}: While ensuring the above two aspects, DIP also has to ensure that the decoder accurately extracts the watermark. Especially when images undergo various image processing operations during network transmission, maintaining accurate traceability is essential. These three optimization objectives can be formalized as follows:

\begin{eqnarray}
    \left\{\begin{matrix} \underset{\mathcal{E}}{min} (\mathcal{L}_{en}(\mathcal{E} (x_{s}  ),x_{s}) ), \; invisibility \ optimization,
 \\\underset{\mathcal{E}}{min} (\mathcal{L}_{adv}(\mathcal{F} (\hat{x}_{s}  ),\mathcal{F} (x_{t}  )) ), \; adversarial \ optimization,
 \\\underset{\mathcal{D}}{min} (\mathcal{L}_{wm}(\mathcal{D} (\hat{x}_{s}  ),w_{id}) ), \; traceability \ optimization,
 \end{matrix}\right.
\end{eqnarray}
where $\mathcal{L}_{en}$, $\mathcal{L}_{adv}$ and $\mathcal{L}_{wm}$ represent the corresponding loss functions, which will be elaborated in the next section.

\section{Methodology}

\begin{figure*}
	\centering
	\includegraphics[width=165mm]{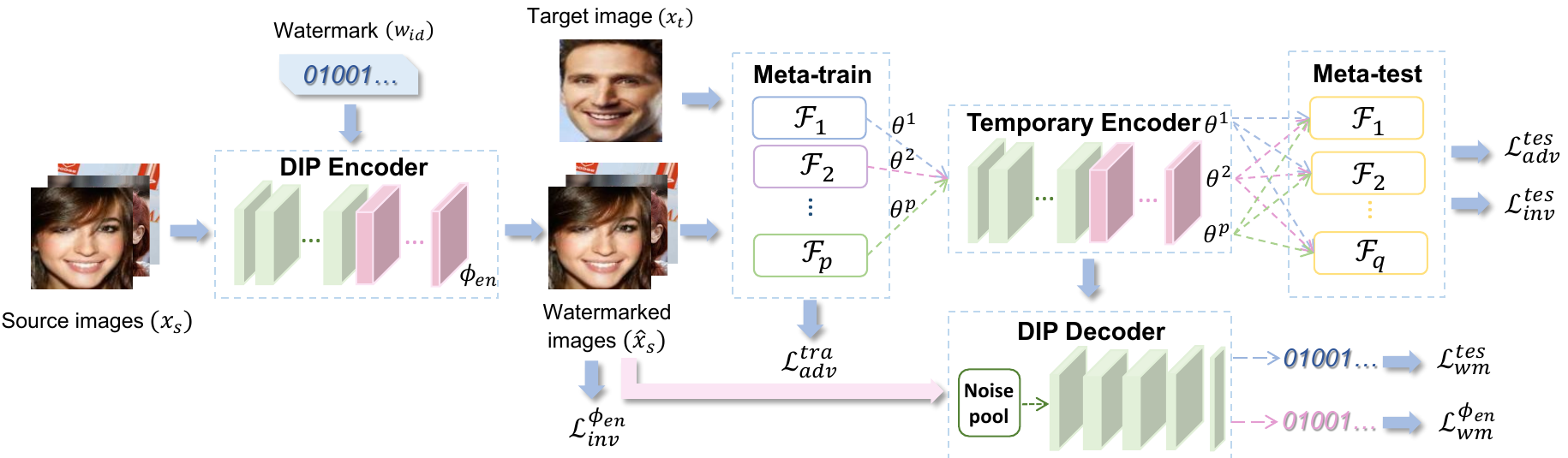}
   
        \caption{The whole pipeline of the proposed  DIP-Watermark. The source image is first encoded with the original parameter encoder to obtain the watermarked image. The watermarked image then enters the meta-learning module for adversarial training using a temporary encoder and undergoes traceability optimization through the noise-pool-based watermark decoder.}	
 \label{fig2}
\end{figure*}

In this section, we provide a detailed explanation of the proposed DIP-Watermark framework. The overall pipeline of DIP-Watermark is illustrated in Fig.~\ref{fig2}, with the complete algorithmic flow presented in Algorithm~\ref{alg}.


\begin{algorithm}[tb]
\caption{DIP-Watermark Training Framework}
\label{alg}
\begin{algorithmic}[1]
\REQUIRE

    $x_{s}$ (training original source image); $w_{id}$ (watermark ID); $\mathcal{E}( \cdot)$ (watermark encoder);
    $\mathcal{D}( \cdot )$ (watermark decoder); 
    $\mathcal{F}_{p}\epsilon \left \{ \mathcal{F}_{1},\mathcal{F}_{2}\dots \mathcal{F}_{P}  \right \}$ (FR models for meta training);
    $\mathcal{F}_{q}\epsilon \left \{ \mathcal{F}_{1},\mathcal{F}_{2}\dots \mathcal{F}_{Q}  \right \}$ (FR models for meta testing);\\
    
    \ENSURE
    best model parameters;

        \STATE  Initialization;    
        \FOR{$i\in  [ 0,maxiter  ]$ }
           \STATE $\hat{x}_{s}  \Leftarrow \mathcal{E}(x_{s}, w_{id}) $;
           
           \FOR{ $p\in P$ }
              \STATE  $\hat{E}_{s}, {E}_{t}  \Leftarrow \mathcal{F}_{p}(\hat{x}_{s }, x_{t}) $;
              \STATE Calculate $\mathcal{L} _{inv}^{\phi_{ en}}$ and $\mathcal{L}_{adv}^{tra}$ with Eq.\ref{ori_enloss}\&\ref{metatrain_advloss};
             \STATE $\theta ^{p}  \longleftarrow \phi_{ en} -\eta \bigtriangledown _{\phi_{ en} } \mathcal{L}_{adv}^{tra}.$
              \FOR{ $q\in Q$ } 
                 \STATE $\hat{x}_{s\_tes}  \Leftarrow \mathcal{E}(x_{s}, w_{id}\mid \theta ^{p}) $;
                  \STATE Calculate $\mathcal{L} _{adv}^{tes}$ and $\mathcal{L} _{inv}^{tes}$ with Eq.\ref{metatest_advloss}\&\ref{metatest_enloss};  
                  \ENDFOR
        
           \ENDFOR
           \STATE Calculate $\mathcal{L} _{inv}^{total}$ and $\mathcal{L} _{adv}^{total}$ with Eq.\ref{entotalloss}\&\ref{advtotalloss};  
           \STATE Calculate $\mathcal{L}_{wm}^{total}$ and $\mathcal{L}_{total}$ with Eq.\ref{wmtotal}\& \ref{total};
           \STATE Update $\mathcal{E}(\cdot)$ and $\mathcal{D}(\cdot)$;
        \ENDFOR
        \RETURN 
        best model parameters;
    \end{algorithmic}
\end{algorithm}

\subsection{Information-Guided Adversarial Attack}

Inspired by the deep watermark model~\cite{jia2021mbrs, zhu2018hidden}, we designed a watermark encoder based on a CNN network structure, processing the carrier image and the watermark sequence in parallel. The carrier first passes through a stacked block Conv-ReLU consisting of the convolutional layer and the ReLU activation. The CBAM~\cite{woo2018cbam} module is used as a channel-wise and spatial-wise attention mechanism to enhance CNN encoding, adaptively recalibrating feature responses to determine the optimal watermark embedding strength. Simultaneously, the Conv-Relu layer processes the watermark sequence in parallel. To seamlessly embed the watermark into the encoded carrier feature map and enhance the watermark's robustness, we introduce redundancy to the watermark using a two-dimensional convolutional layer with a stride of 2, expanding it to the same dimension as the carrier. The CBAM module is similarly applied to enhance watermark feature encoding. Finally, the watermark sequence is concatenated with the carrier feature maps, followed by two-dimensional convolutional fusion, resulting in the watermarked image.


The watermark encoder $\mathcal{E}(\cdot)$ receives the original face image $x_{s}$ along with the watermark $w_{id}$  representing its identity and encodes them under initialized parameters $\phi_{ en}$:

\begin{eqnarray}\label{en}
\hat{x} _{s}=\mathcal{E}(x_{s},w_{id}\mid \phi_{ en} ).
\end{eqnarray}
To enhance the generalization capability of the  DIP-Watermark encoder and decoder and avoid overfitting, we generate a random watermark for each batch during training process. In practical use, the platform registers an identity-specific watermark for each user, and the watermark information can be customized by the platform according to the users. To limit the impact of watermark encoding on the carrier, we use MSE loss $\mathcal{L} _{MSE}$ to define the image invisibility loss $\mathcal{L} _{inv}^{\phi_{ en}}$:


\begin{eqnarray}\label{ori_enloss}
\mathcal{L} _{inv}^{\phi_{ en}}=\mathcal{L} _{MSE}(x_{s}, \hat{x} _{s}).
\end{eqnarray}

To make the traceable watermark adversarial against FR models, we introduce adversarial constraints to optimize the encoder. FR model attacks can be categorized into two types: untargeted (dodging) and targeted (impersonation) attacks. dodging attacks aim to reduce the similarity of images of the same identity to attack the FR model, while impersonation attacks aim to enhance feature similarity between target and source images of different identities~\cite{yin2021adv}, thus disrupting the decision of the FR model. Dodging attacks hinder the stable training of the watermark model, easily falling into a local optimum of watermarked image quality optimization in the early stages of training. Impersonation attacks can seamlessly integrate into the end-to-end optimization process of the watermark model, unifying the multi-task optimization direction of the watermark, and disrupting the feature reconstruction of the FR model. Therefore, we focus on impersonation attacks as in~\cite{yin2021adv,Li_2023_CVPR,liu2023diffprotect}.


We map the identifiable features of the source image to the target identity domain by optimizing the embedding of watermark information. we first obtain recognizable feature embeddings $E_{\ast}$ from the FR model for both the watermarked image and the target image, as follows:


\begin{eqnarray}\label{fr}
\hat{E}_{s},E_{t} =\mathcal{F}(\hat{x}_{s},x_{t}).
\end{eqnarray}
These features are then used to calculate the adversarial loss $\mathcal{L}_{adv}$ for targeted attacks. We employ cosine similarity loss~\cite{deng2019arcface} to calculate $\mathcal{L}_{adv}$:

\begin{eqnarray}\label{advverloss}
\mathcal{L} _{adv} =1-cos(\hat{E}_{s},E_{t} ).
\end{eqnarray}


Existing research has demonstrated that multitask learning improves adversarial robustness while maintaining single-task performance~\cite{mao2020multitask}. This further motivates the expansion of the output dimensions of the watermarking model by incorporating adversarial attack tasks.



\subsection{Muti-Task Compatible Meta-Optimization}

\begin{figure}
	\centering
	\includegraphics[width=80mm]{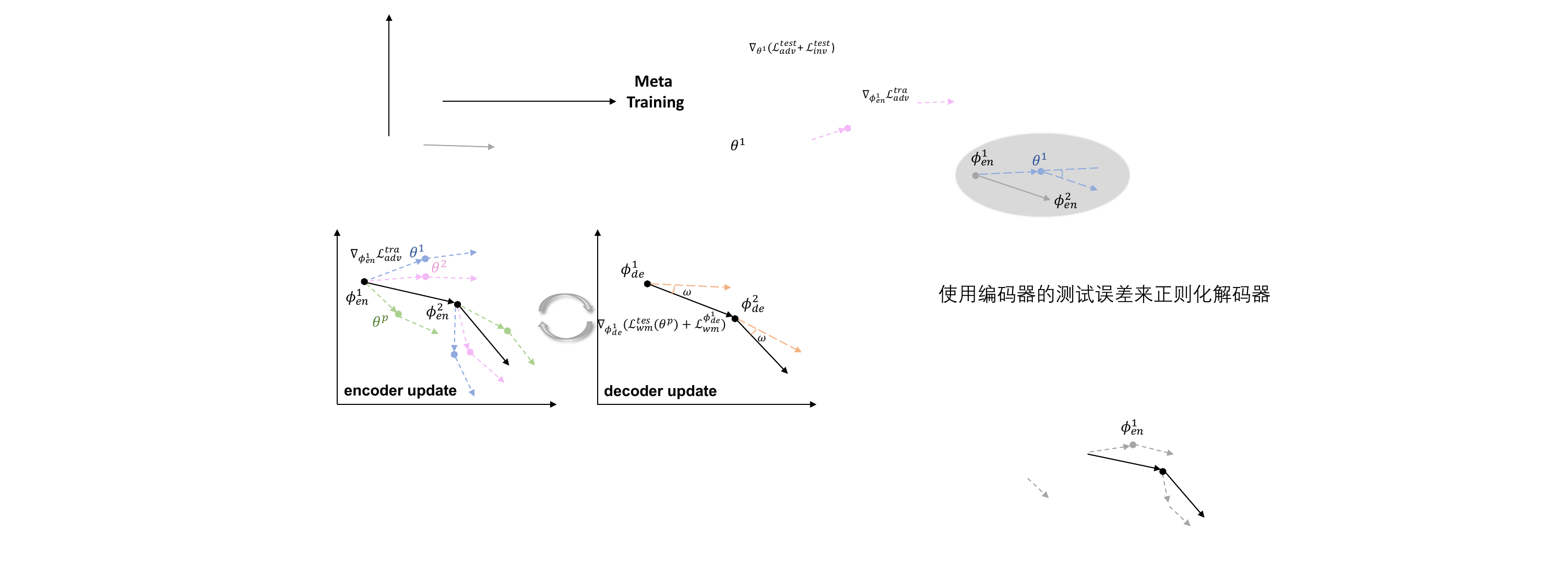}
   
        \caption{Illustration of encoder and decoder co-optimization. The encoder optimizes towards a general direction through meta-optimization, while the decoder simultaneously adjusts its optimization direction based on the temporary parameters of the encoder.}	
 \label{meta}
\end{figure}

The  DIP-Watermark encoder is associated with two distinct sub-tasks: \textit{1) encoding the watermark recoverably and invisibly}; \textit{2) adversarial encoding the watermark to deceive FR models}. Simultaneously, the decoder needs to accurately extract the watermark.  DIP-Watermark are optimized end-to-end to search for the global optimal solution. However, during the watermark model training process, there are optimization conflicts between different tasks.  Enhancing the watermark adversariality on a single FR model or on a simple ensemble of FR models would lead the encoder to blindly enhance the watermark embedding strength, thus hindering the optimization of watermark invisibility, introducing more noticeable perceptual artifacts into the watermarked image. More importantly, such adversariality enhancement cannot be effectively transferred to black-box models. Therefore, to enhance the transferability and mitigate this optimization conflict, we introduce a collaborative meta-optimization strategy compatible with multi-task. This strategy builds upon Model-Agnostic Meta-Learning (MAML) method~\cite{finn2017model}. MAML encodes more representative shared adversarial features from the carrier by combining different known models, thus quickly adapting to unknown target models. We seamlessly integrate adversarial transferability optimization into other subtasks of the encoder to aid the overall optimization of the watermark model.The schematic diagram of model parameter updates in the overall optimization is shown in Fig.~\ref{meta}.

In the meta-optimization strategy, the known target victim models are divided into meta-training models and meta-testing models. Temporary model parameters are calculated using the loss obtained on the meta-training models, and the temporary parameters are utilized to compute the testing loss on the meta-testing models. We select $P$ meta-training models and $Q$ meta-testing models, initially calculating the adversarial loss under the meta-training models with the initial encoder parameters $\phi_{ en}$:

\begin{eqnarray}\label{metatrain_advloss}
\mathcal{L}_{adv}^{tra}=\sum_{p=1}^{P} \mathcal{L}_{adv}(\mathcal{F}_{p}(\hat{x}_{s}),\mathcal{F}_{p}(x_{t})\mid \phi_{ en} ),
\end{eqnarray}
where $\mathcal{F}_{p}$ represents the $p$-th meta-training FR model, and $\mathcal{L}_{adv}^{tra}$ represents the obtained meta-training adversarial loss.

For each meta-training model, we calculate the gradient of the adversarial loss with respect to the encoder parameters based on the model, and update the temporary parameters $\theta ^{p}$ of the current encoder according to this gradient:


\begin{eqnarray}\label{fast}
\theta ^{p}  \longleftarrow \phi_{ en} -\eta \bigtriangledown _{\phi_{ en} } \mathcal{L}_{adv}^{tra}.
\end{eqnarray}

The encoder utilizes temporary parameters to generate the watermarked image $\hat{x} _{s\_tes}$:

\begin{eqnarray}\label{test_img}
\hat{x} _{s\_tes} = \mathcal{E}(x_{s},w_{id}\mid\theta ^{p}).
\end{eqnarray}
Temporary watermarked images are used to compute test errors on meta-testing models, assessing the meta-performance across $P$ tasks and refining the original optimization direction. The meta-testing adversarial loss $\mathcal{L} _{meta}^{tes}$ is defined as:

\begin{eqnarray}\label{metatest_advloss}
\mathcal{L} _{adv}^{tes}=  \sum_{p=1}^{P}\sum_{q=1}^{Q} \mathcal{L} _{adv}(\mathcal{F}_{q}(\hat{x}_{s\_tes}),\mathcal{F}_{q}(x_{t})\mid \theta ^{p}).
\end{eqnarray}

To ensure compatibility with the invisibility optimization task and further constrain image degradation due to adversarial enhancement, we concurrently calculate the invisibility loss $\mathcal{L} _{inv}^{tes}$ based on $\hat{x} _{s\_tes}$ under temporary parameters, unifying the gradient optimization direction:
\begin{eqnarray}\label{metatest_enloss}
\mathcal{L} _{inv}^{tes}=\frac{1}{P Q}\sum_{p=1}^{P}\sum_{q=1}^{Q}\mathcal{L} _{MSE}(\hat{x}_{s\_tes},x_{s}\mid \theta ^{p}),
\end{eqnarray}
the total invisibility loss $\mathcal{L} _{inv}^{total}$ and total adversarial loss $\mathcal{L} _{adv}^{total}$ during end-to-end updates are calculated as follows:
\begin{eqnarray}\label{entotalloss}
\mathcal{L} _{inv}^{total}=  (\mathcal{L}_{inv}^{\phi_{en}} + \mathcal{L} _{inv}^{tes})/2.
\end{eqnarray}

\begin{eqnarray}\label{advtotalloss}
\mathcal{L} _{adv}^{total}=  \mathcal{L}_{adv}^{tra} + \mathcal{L} _{adv}^{tes}.
\end{eqnarray}

\subsection{Regularization-Enhanced Information Decoding}

The online network processes images uploaded to the platform to conform to the standards for image file storage and transmission. This complex image processing often significantly alters watermarked images, increasing the difficulty of watermark extraction. Therefore, we design a joint noise pool $NP(\cdot)$ that includes differentiable JPEG compression ($QF\!\!=\!\!50$) and Gaussian noise ($Var=0.003$) to enhance the decoder robustness. Watermarked images under both temporary and initial parameters are uniformly processed through the noise pool before decoding. We initially calculate the information loss of the watermarked image obtained under the original encoder parameters using Binary Cross-Entropy (BCE) loss $\mathcal{L} _{BCE}$, as follows:

\begin{eqnarray}\label{wmori}\begin{aligned}
\mathcal{L}_{wm}^{\phi_{en}}  =\mathcal{L} _{BCE}( w_{id},\mathcal{D}(NP(\mathcal{E}(x_{s},w_{id}\mid \phi_{ en} )))).
\end{aligned}
\end{eqnarray}

The encoder and decoder interact synergistically during co-optimization, serving as mutual priors, as illustrated in Fig.~\ref{meta}. To enhance cross-model gradient compatibility, we introduce an information regularization module. This module calculates information loss using the latest temporary update of the watermarked image, allowing it to further refine the optimization direction of the decoder based on the encoder:

\begin{eqnarray}\label{wmtes}\begin{aligned}
\mathcal{L}_{wm}^{tes}  =\mathcal{L} _{BCE}( w_{id},\mathcal{D}(NP(\mathcal{E}(x_{s},w_{id}\mid\theta ^{P})))).
\end{aligned}
\end{eqnarray}
The overall watermark information loss is calculated by:


\begin{eqnarray}\label{wmtotal}\begin{aligned}
\mathcal{L}_{wm}^{total}  =\mathcal{L}_{wm}^{\phi_{en}} + \mathcal{L}_{wm}^{tes}.
\end{aligned}
\end{eqnarray}



The outer loop optimization of the model is guided by three types of losses: adversarial loss, invisibility loss, and watermark information loss.These losses collectively serve as optimization objectives to simultaneously update the encoder and decoder, searching for the global optimum. The total loss calculation for DIP-Watermark is as follows:


\begin{eqnarray}\label{total}\begin{aligned}
\mathcal{L}_{DIP}  =\lambda _{adv}\mathcal{L} _{adv}^{total}+\lambda _{inv}\mathcal{L} _{en}^{total}+\lambda _{wm}\mathcal{L}_{wm}^{total},
\end{aligned}
\end{eqnarray}
where $\lambda _{adv}$, $\lambda _{en}$ and $\lambda _{wm}$ are positive hyperparameters, and we empirically set them to 100, 0.05 and 0.05, respectively. To balance the significantly different loss values among various tasks and ensure stable training, we set different orders of magnitude for loss weights. During the initial optimization phase, watermark embedding results in severe degradation of the watermark image. At this stage, the invisibility loss measured at the pixel level is considerably larger compared to the adversarial loss. Thus, we address this by assigning smaller weights to the former. Furthermore, we maintain consistency in the orders of magnitude between information loss and image loss weights. This is because, during the initial optimization, the watermark image resembles meaningless noise, and the decoder does not need to learn watermark information from such noise distribution. As the invisibility loss rapidly decreases to the same order of magnitude as the watermark information loss, the watermark image exhibits a learnable pattern distribution, allowing for more balanced and stable updates to the decoder. The impact of weight coefficient settings on performance  is demonstrated in the ablation study section.

\section{Experiments}
\subsection{Experimental Settings}
\subsubsection{Datasets} We evaluate  DIP-Watermark on two commonly used large face image datasets: LFW~\cite{huang2008labeled} and CelebA~\cite{liu2015deep}. The LFW dataset consists of 13,233 facial images belonging to 5,749 identities. CelebA comprises over 200,000 facial images. We randomly selected 3,000 images from each dataset for watermark model training and 1,000 images for testing. The watermark message is 50 bits, sufficient to meet the requirements for representing identity in the real world. To meet the input size requirements of FR models, all images are initially processed through MTCNN~\cite{7553523} to extract facial regions and sampled to a size of $112\times 112$. 

\subsubsection{Implementation details}  Our DIP-Watermark is implemented by PyTorch and executed on NVIDIA RTX 3090. The entire training process of DIP-Watermark spans 2500 epochs with a batch size of 32. We empirically adjust the Adam optimizer with an initial learning rate of 0.00005 for stable training. We select various offline and online commercial victim FR models to evaluate the black-box transferability of  DIP-Watermark,including: \textit{1) Offline models}: IR152, IRSE50, MobileFace~\cite{deng2019arcface}, FaceNet~\cite{10.1145/2976749.2978392}, ArcFace~\cite{deng2019arcface} based on IResNet100~\cite{duta2021improved} and CosFace~\cite{wang2018cosface} based on IResNet100. \textit{2) Online models}: Face++~\cite{MEGVII} and Aliyun~\cite{Aliyun}. To demonstrate the black-box transferability of our method, we alternately selected three models from them as white-box models for integrated training and sequentially obtained black-box test results on the remaining models as in~\cite{yin2021adv,komkov2021advhat,hu2022protecting}, with the parameters set as $P=2$ and $Q=1$.


\subsubsection{Baselines} 


 DIP-Watermark is the first watermarking method designed for FR models with both adversariality and traceability, and there is no similar work before. Thus, we first compare its adversariality with adversarial attack methods: \textit{1) Patch-based adversarial attack methods}: Adv-Makeup~\cite{yin2021adv}, AdvHat~\cite{komkov2021advhat}; \textit{2) Perturbation-based adversarial attack methods}: PGD~\cite{madry2017towards}, FGSM~\cite{goodfellow2014explaining}. Simultaneously, we straightforwardly combine adversarial adversarial attack methods with watermarking methods Hidden~\cite{zhu2018hidden} and FakeTagger~\cite{wang2021faketagger}. We embed watermarks directly into adversarial examples generated by PGD and FGSM, By comparing adversariality and traceability, we demonstrate the advantages of DIP-Watermark, which only requires embedding a single watermark.

\begin{table*}[t]
 
	\renewcommand\arraystretch{1.2}  
\resizebox{\hsize}{!}{
 
       \begin{tabular}{c|c|c|cc|cc|cc|cc|cc|cc}
       \Xhline{1px}
           \multirow{2}{*}{\textbf{Datasets}} &\multirow{2}{*}{\textbf{Type}} & \multirow{2}{*}{\textbf{Method}}&\multicolumn{2}{c}{\textbf{IR152}} &\multicolumn{2}{c}{\textbf{IRSE50}}&\multicolumn{2}{c}{\textbf{IResNet100-Arc}} &\multicolumn{2}{c}{\textbf{MobileFace}} & \multicolumn{2}{c}{\textbf{FaceNet}} &\multicolumn{2}{c}{\textbf{IResNet100-Cos}} 
            \\ \cline{4-15}

			&&& ASR&ACC& ASR&ACC& ASR&ACC& ASR&ACC& ASR&ACC& ASR&ACC \\
            \hline
		\multirow{9}{*}{LFW}&\multirow{2}{*}{Patch-based}
     &Adv-Makeup & 0.026 & N/A & 0.236 &N/A&0.026 &N/A &0.289 &N/A & 0.026 &N/A  &0.000&N/A \\
		& &AdvHat&0.017  &N/A &0.187 &N/A &0.000 &N/A &0.215  &N/A &0.008 &N/A  &0.000  &N/A \\
			\cline{2-15}
             &\multirow{2}{*}{Perturbation-based}
             &PGD &\underline{0.311} &N/A &\underline{0.788} &N/A&\underline{0.367}  &N/A & \textbf{0.755} &N/A &\underline{0.155} &N/A  &\underline{0.266} & N/A \\
			 &&FGSM &0.177 & N/A & 0.588 &N/A & 0.111 &N/A &0.622&N/A & 0.067 &N/A & 0.044 &N/A \\
			
              \cline{2-15}
               &\multirow{4}{*}{Combination-based}
               &PGD$+$Hidden &0.182 &0.988 &0.501 &0.988&0.065  &0.984 &0.433  &0.988 &0.062 &0.984  &0.037 &0.967  \\
			 &&FGSM$+$Hidden &0.109 &0.988  &0.310  &0.988&0.011  &0.983 &0.299 &0.988  &0.011  &0.983 &0.015 &0.983 \\
			
     \cline{3-15}
    &&PGD$+$FakeTagger &0.265 &\underline{0.991} &0.532 &\underline{0.989}& 0.050 &\underline{0.995} &0.483 &\underline{0.989} &0.040 &\underline{0.991}  &0.030 & 0.986 \\
			 &&FGSM$+$FakeTagger&0.112 &0.986  &0.350  &0.986 & 0.020 &0.988 &0.391 &0.986 & 0.010 &0.988 & 0.020 &\underline{0.988} \\
			
              \cline{2-15}
             & Watermark-based &\textbf{ DIP-Watermark (ours)} &\textbf{0.326} &\textbf{0.992} & \textbf{0.810} &\textbf{0.997}&\textbf{0.381} &\textbf{0.996} &\underline{0.716} &\textbf{0.997} & \textbf{0.643} &\textbf{0.994}  &\textbf{0.355} &\textbf{0.996}\\
			\hline
   \hline
   \multirow{9}{*}{CelebA}&\multirow{2}{*}{Patch-based}&Adv-Makeup & 0.087 & N/A &0.193 &N/A  & 0.017 &N/A&0.000 &N/A &0.421 &N/A  &0.00 &N/A\\
    &&AdvHat &0.017 &N/A  &0.072 &N/A&0.000 &N/A &0.071  &N/A     &0.000  &N/A &0.000  &N/A\\
    \cline{2-15}
    &\multirow{2}{*}{Perturbation-based}&PGD & \underline{0.400} &N/A &\textbf {0.660}  &N/A  & \textbf{0.450} &N/A&\textbf{0.720} &N/A &\underline{ 0.270 }&N/A &\underline{0.378} &N/A\\
    &&FGSM &0.180 &N/A& 0.420 &N/A& 0.220 &N/A &0.600 & N/A & 0.100 &N/A  &0.180 & N/A\\
	
 \cline{2-15}

    &\multirow{4}{*}{Combination-based}
    &PGD$+$Hidden &0.132 &0.978 &0.291 &0.983&0.141 &0.957 &0.181 &0.979 &0.143 &0.960  &0.090 &0.960  \\
			 &&FGSM$+$Hidden&0.122 &0.975 &0.171 &0.975&0.080 &0.968 &0.100 &0.958 &0.030 &0.968  &0.000 &0.966  \\
			 
             \cline{3-15}
              & &PGD$+$FakeTagger &0.376 &\underline{0.986} &0.380 &\underline{0.993} & 0.240 &0.988 &0.364 &\underline{0.993} &0.171 &0.984  &0.290 & 0.986 \\
			 &&FGSM$+$FakeTagger&0.311 &\underline{0.986} & 0.240 &0.992 & 0.111 &\underline{0.989} &0.281 &0.992 & 0.060 &\underline{0.992}  & 0.080 &\underline{0.987} \\
			
     \cline{2-15}
 &Watermark-based &\textbf{ DIP-Watermark (ours)}&\textbf{0.433} &\textbf{0.998} &\underline{0.635} &\textbf{0.998}&\underline{0.375} &\textbf{0.995} &\underline{0.679}  &\textbf{0.997} &\textbf{0.318}  &\textbf{0.995}  &\textbf{0.393} &\textbf{0.998} \\
\Xhline{1px}
	\end{tabular}}
	\caption{ASR and ACC results of black-box impersonation attack on the LFW and CelebA datasets. $N/A$ indicates that the method cannot achieve this effect and lacks supporting data. The best results are emphasized in bold, the second-best results are underscored.}
	\label{table1}

\end{table*}

\begin{table}[t]

	\renewcommand\arraystretch{1.2}  
\resizebox{\hsize}{!}{
 
       \begin{tabular}{c|c|cc|cc}
       \Xhline{1px}
           \multirow{2}{*}{\textbf{Type}} & \multirow{2}{*}{\textbf{Method}}&\multicolumn{2}{c}{\textbf{Face++}} &\multicolumn{2}{c}{\textbf{Aliyun}}
            \\ \cline{3-6}

			&& ACS &ACC & ACS &ACC \\
            \hline
			\multirow{2}{*}{Patch-based}&Adv-makeup &52.320  &N/A &6.603 &N/A  \\
			&Adv-Hat&48.030  &N/A &3.493 &N/A  \\
		 
			\hline
             \multirow{2}{*}{Perturbation-based}&PGD&\underline{65.000}  &N/A &\underline{41.517} &N/A  \\
			 &FGSM&58.490  &N/A &27.780 &N/A  \\
			
              \hline
               \multirow{4}{*}{Combination-based}&PGD$+$Hidden &49.748 &0.983  &27.067 &0.993    \\
			 &FGSM$+$Hidden &45.889  &0.975 &22.095 &0.975 \\
			
     \cline{2-6}
    &PGD$+$FakeTagger &54.370 &\underline{0.993}  &33.572 &\underline{0.993} \\
			 &FGSM$+$FakeTagger &46.325  &0.992 &22.537 &0.992 \\
			
              \hline
             Watermark-based &\textbf{ DIP-Watermark (ours)} &\textbf{65.491}  &\textbf{0.998} &\textbf{43.945} &\textbf{0.998} \\
			\hline

\Xhline{1px}

	\end{tabular}}
\caption{ACS and ACC results on online commercial FR systems.}
	\label{online}

\end{table}

\subsubsection{Metrics} We use SSIM~\cite{Wang_Bovik_Sheikh_Simoncelli_2004} to evaluate the quality of images. Accuracy $Acc$ is used to evaluate the watermark recovery accuracy of the watermarked image, which is calculated as follows:
\begin{eqnarray}\label{ACC}
Acc=1-\frac{1}{L}\sum_{l=1}^{L}\left | w _{id}-\hat{w} _{id}   \right |,   
\end{eqnarray}
where $L$ denotes the total length of the watermark information and $l$ denotes the index of the watermark bit.

We adopt Attack Success Rate (ASR)~\cite{deb2020advfaces} for impersonation attack to evaluate adversariality, which is calculated through: 
\begin{eqnarray}\label{ASR}
ASR=\frac{\sum_{i}^{N}cos[\mathcal{F}(x_{t}^{i}),\mathcal{F}(\hat{x} _{s}^{i})]>\tau }{N},
\end{eqnarray}
where $i$ represents the image index, $N$ represents the total number of test images, and $\tau$ represents the similarity threshold. The value of $\tau$ will be set as the threshold at 0.01 FAR (False Acceptance Rate) for each victim FR model as most face recognition works do, i.e., IR152 (0.167), IRSE50 (0.241), MobileFace (0.302), FaceNet (0.409), IResNet100-Arc (0.280) and IResNet100-Cos (0.280)~\cite{yin2021adv}.

\subsection{Comparison with Baselines}


To meet the challenges of the practical scenarios, where users are entirely unaware of the underlying structure and the specific training schemes of the FR models used by third-party databases, We benchmark against prior work~\cite{yin2021adv,Li_2023_CVPR,liu2023diffprotect}, focusing on evaluating the defense method's performance against black-box models. For perturbation-based adversarial attack methods, to ensure the generated adversarial images do not affect social usability, we control the average SSIM value of all adversarial images generated by different methods to be 0.9 $\pm $ 0.03 in our experiments for fair comparison. All adversarial attack methods are integrated attacks conducted on same victim models as DIP-Watermark. 

As shown in Table~\ref{table1},  DIP-Watermark exhibits excellent adversariality and traceability simultaneously. It achieves optimal ASR on almost evaluated black-box FR models, surpassing adversarial example methods that only possess adversariality. Patch-based adversarial attacks generally exhibit poor black-box transferability, particularly with minimal adversarial impact on CosFace and ArcFace models. Notably, Patch-based adversarial attack methods directly alter low-level pixel information, affecting image appearance and social usability. Perturbation-based methods exhibit relatively better black-box transferability; however, all methods consistently demonstrate lower black-box ASR on FaceNet, CosFace, and ArcFace, which can be attributed to the fact that, unlike feature-matching-based FR models, FaceNet with InceptionResnetV1 as the backbone is a classification-based FR model, and the IRSeNet100 shares a similar ResNet variant with it, making transfer attacks more challenging. Nevertheless, DIP-Watermark achieves nearly optimal black-box ASR, especially surpassing perturbation-based methods by up to $50\%$ on FaceNet. While ensuring excellent black-box ASR, DIP-Watermark also maintains superior traceability accuracy. Combined-based methods significantly compromise adversariality after watermark embedding, and embedding watermark information after each image iteration for adversarial example generation drastically reduces computational efficiency. In contrast, DIP-Watermark requires embedding only one robust watermark, with an embedding response time of merely 0.07s, while achieving outstanding black-box ASR and over $99\%$ watermark recovery accuracy.

To assess the practical feasibility of the methods, we employed two commercial online FR APIs for face verification, including Face++~\cite{MEGVII} and Aliyun~\cite{Aliyun}. We upload the adversarial images generated by the different methods with the target facial images to the platforms, and the white-box models used for training are MobileFace, FaceNet, and CosFace. The platforms provided confidence scores indicating whether these two images belong to the same identity. We report the average confidence score (ACS) for 100 images from the LFW dataset. As shown in Table~\ref{online},  DIP-Watermark achieved the highest ACS while maintaining excellent traceability accuracy. This confirms the practical deployability of DIP-Watermark.

\begin{table*}[t]
	 
	\renewcommand\arraystretch{1.2}  
\resizebox{\hsize}{!}{
 
       \begin{tabular}{c|c|c|c|cc|cc|cc|cc|cc|cc}

       \Xhline{1px}
           \multirow{2}{*}{\textbf{Processing}}&\multirow{2}{*}{\textbf{Parameters}}  &\multirow{2}{*}{\textbf{Type}} & \multirow{2}{*}{\textbf{Method}}&\multicolumn{2}{c}{\textbf{IR152}} &\multicolumn{2}{c}{\textbf{IRSE50}}&\multicolumn{2}{c}{\textbf{IResNet100-Arc}}&\multicolumn{2}{c}{\textbf{MobileFace}} & \multicolumn{2}{c}{\textbf{FaceNet}}  &\multicolumn{2}{c}{\textbf{IResNet100-Cos}} 
            \\ \cline{5-16}

	&&&& ASR &ACC & ASR &ACC & ASR &ACC & ASR &ACC & ASR &ACC & ASR &ACC\\
    \hline
    
	\multirow{9}{*}{\makecell[c]{Gaussian\\ noise}} &\multirow{9}{*}{Var=0.003} 
    &\multirow{3}{*}{Patch-based}
    &Adv-Makeup &0.026 &N/A  &0.263 &N/A &0.026 &N/A &0.289  &N/A  &0.026 &N/A   &0.000  &N/A  \\
	&&&AdvHat&0.017 &N/A  &0.008 &N/A &0.000 &N/A &0.071  &N/A  &0.000 &N/A   &0.000  &N/A  \\
	\cline{3-16}
   
    &&\multirow{2}{*}{Perturbation-based}&PGD &0.191 &N/A &\underline{0.465} &N/A &0.011 &N/A&\underline{0.372} &N/A &0.015 &N/A   &0.011 &N/A  \\
	& &&FGSM &0.150  &N/A &0.292 & N/A &0.017 &N/A &0.305  &N/A &0.000 &N/A  &0.000 &N/A \\
    \cline{3-16}

    &&\multirow{4}{*}{Combination-based}&PGD$+$Hidden &0.211 &0.629 &0.424 &0.636&\underline{0.055} &0.632 &0.365 &0.629 &\underline{0.033} &0.632  &0.016 &0.632  \\
	&&&FGSM$+$Hidden&0.113 &0.652 &0.145 &0.633 &0.000 &0.636 &0.312 &0.633 &0.011 &0.636  &0.000 &0.641  \\
	
    \cline{4-16}
    && &PGD$+$FakeTagger &\underline{0.220} &\underline{0.982}  &0.410 &\underline{0.971} &0.023  &0.961&0.341  &\underline{0.971} &0.017 &\underline{0.982}  &0.022 &0.961  \\
	&&&FGSM$+$FakeTagger&0.103 &0.975 &0.312  &0.970 & 0.013 &\underline{0.963}&0.300 &0.963 & 0.011 &0.970  & \underline{0.024} &\underline{0.963} \\
    \cline{3-16}
    
    &&Watermark-based 
    &\textbf{ DIP-Watermark (ours)} &\textbf{0.265} &\textbf{0.975} &\textbf{0.725}  &\textbf{0.987} &\textbf{0.308}  &\textbf{0.979}&\textbf{0.695} &\textbf{0.987} &\textbf{0.563}  &\textbf{0.975}   &\textbf{0.311}   &\textbf{0.985} \\
	\hline
   \hline
   
   \multirow{9}{*}{JPEG} &\multirow{9}{*}{QF=30}
   
   &\multirow{2}{*}{Patch-based}
   &Adv-Makeup &0.000 &N/A  &0.052 &N/A&0.000 &N/A &0.131  &N/A  &0.000 &N/A   &0.000  &N/A \\
   &&&AdvHat &0.008 &N/A  &0.008 &N/A &0.000 &N/A &0.035  &N/A  &0.000 &N/A  &0.000  &N/A \\
   \cline{3-16}

   &&\multirow{2}{*}{Gradient-based}
   &PGD  &0.244 &N/A &\underline{0.523} &N/A&\underline{0.051} &N/A &\underline{0.480} &N/A &0.026 &N/A  &0.033 &N/A\\
   &&&FGSM &0.205 &N/A &0.371 &N/A&0.013 &N/A &0.375 &N/A &0.022 &N/A  &0.011 &N/A\\

   \cline{3-16}

   &&\multirow{4}{*}{Combination-based}
   &PGD$+$Hidden  &\underline{0.261} &0.667 &0.513 &0.654&0.032 &0.662 &0.412 &0.654 &0.030 &0.672  &\underline{0.045} &0.654  \\
   &&&FGSM$+$Hidden &0.119 &0.686 &0.280 &0.686&0.000 &0.673  &0.295 &0.692 &\underline{0.035} &0.655 &0.016 &0.672  \\
  
   \cline{4-16}
   &&&PGD$+$FakeTagger &0.260 &0.886 &0.463 &0.892&0.040  &0.883 &0.440 &\underline{0.898} &0.021 &0.883  &0.029 &\underline{0.898} \\
   &&&FGSM$+$FakeTagger&0.119 &\underline{0.887} & 0.310 &\underline{0.896}&0.016  &\underline{0.883} &0.360 &0.887 &0.022  &\underline{0.893}  &0.011  &0.883 \\
   \cline{3-16}
    
    &&Watermark-based 
    &\textbf{ DIP-Watermark (ours)}&\textbf{0.272} &\textbf{0.967} &\textbf{0.655} &\textbf{0.958} &\textbf{0.286} &\textbf{0.983} &\textbf{0.736}  &\textbf{0.967} &\textbf{0.587}  &\textbf{0.983}  &\textbf{0.398}  &\textbf{0.980}  \\
    \hline
    \hline

   \multirow{9}{*}{Resize}&\multirow{9}{*}{Size=$1/2$}
   &\multirow{2}{*}{Patch-based}&Adv-Makeup &0.026 &N/A  &0.236 &N/A&0.026 &N/A  &0.289  &N/A   &0.026 &N/A &0.000  &N/A \\
   &&&AdvHat&0.008  &N/A &0.008 &N/A  &0.000 &N/A &0.003 &N/A &0.000 &N/A  &0.000  &N/A \\
   \cline{3-16}
    
   &&\multirow{2}{*}{Perturbation-based}
   &PGD &0.233 &N/A  &0.380 &N/A &\underline{0.039} &N/A &0.433  &N/A  &0.017 &N/A   &\underline{0.028}  &N/A \\
   &&&FGSM &0.200 &N/A &0.310 &N/A&0.000 &N/A  &0.351 &N/A &0.016 &N/A  &0.000  &N/A \\
   \cline{3-16}
   
    &&\multirow{4}{*}{Combination-based}
    &PGD$+$Hidden &\underline{0.241} &0.685  &\underline{0.411} &0.691 &0.023 &0.641 &\underline{0.444}  &0.685  &\underline{0.022} &0.663   &0.023  &0.641   \\
	&&&FGSM$+$Hidden &0.109 &0.714  &0.312 &0.762 &0.000 &0.721 &0.333  &0.754  &0.011 &0. 721  &0.000  &0.736   \\
	
    \cline{4-16}
    &&&PGD$+$FakeTagger  &0.203 &0.987  &0.336 &0.977 &0.025 &\underline{0.988} &0.425  &0.987  &0.011 &0.979   &0.016  &0.982 \\
	&&&FGSM$+$FakeTagger&0.137 &\underline{0.987}  &0.246 &\underline{0.986}&0.000 &0.982 &0.361  &\underline{0.987}  &0.011 &\underline{0.986}   &0.000  &\underline{0.986}  \\
	
    \cline{3-16}
    
    &&Watermark-based 
    &\textbf{ DIP-Watermark (ours)}&\textbf{0.255} &\textbf{0.990}  &\textbf{0.584} &\textbf{0.990} &\textbf{0.211} &\textbf{0.989 } & \textbf{0.688} &\textbf{0.990 } &\textbf{0.408} &\textbf{0.993}  &\textbf{0.253}   & \textbf{0.988}  \\
    \hline
    \hline

       \multirow{9}{*}{\makecell[c]{Gaussian\\ blur}}&\multirow{9}{*}{\makecell[c]{Kernel=3\\SD=30}}
   &\multirow{2}{*}{Patch-based}&Adv-Makeup &0.057 &N/A  &0.192 &N/A&0.057 &N/A  &0.326  &N/A   &0.010 &N/A &0.002  &N/A \\
   &&&AdvHat&0.000  &N/A &0.009 &N/A  &0.000 &N/A &0.047 &N/A &0.000 &N/A  &0.000  &N/A \\
   \cline{3-16}
    
   &&\multirow{2}{*}{Perturbation-based}
   &PGD &0.259 &N/A  &\underline{0.440} &N/A &\underline{0.060} &N/A &\underline{0.481}  &N/A  &\underline{0.042} &N/A   &\underline{0.036}  &N/A \\
   &&&FGSM &0.196 &N/A &0.320 &N/A&0.022 &N/A  &0.361 &N/A &0.034 &N/A  &0.011  &N/A \\
   \cline{3-16}
   
    &&\multirow{4}{*}{Combination-based}
    &PGD$+$Hidden &0.251 &0.664  &0.403 &0.675 &0.030 &0.664 &0.452  &0.686  &0.033 &0.683   &0.021  &0.672   \\
	&&&FGSM$+$Hidden &0.176 &0.689  &0.306 &0.689 &0.011 &0.672 &0.320  &0.679  &0.010 &0. 689  &0.000  &0.689   \\
	
    \cline{4-16}
    &&&PGD$+$FakeTagger  &\underline{0.272} &\underline{0.986}  &0.387 &0.983 &0.051 &0.976 &0.416  &\underline{0.979}  &0.022 &0.968   &0.020  &\underline{0.983}  \\
	&&&FGSM$+$FakeTagger&0.183 &0.981  &0.342 &\underline{0.985}&0.000 &\underline{0.980} &0.375  &0.973  &0.010 &\underline{0.977}   &0.023  &0.972  \\
	
    \cline{3-16}
    
    &&Watermark-based 
    &\textbf{DIP-Watermark (ours)}&\textbf{0.296} &\textbf{0.991}  &\textbf{0.733} &\textbf{0.987} &\textbf{0.265} &\textbf{0.987 } & \textbf{0.668} &\textbf{0.992 } &\textbf{0.527} &\textbf{0.988}  &\textbf{0.211}   & \textbf{0.986}  \\
    
   \Xhline{1px}
	\end{tabular}}
\caption{ASR and ACC results after various image processing operations. The testing dataset used is the LFW dataset. }
	\label{table2}

\end{table*}

\begin{figure}
	\centering
	\includegraphics[width=86mm]{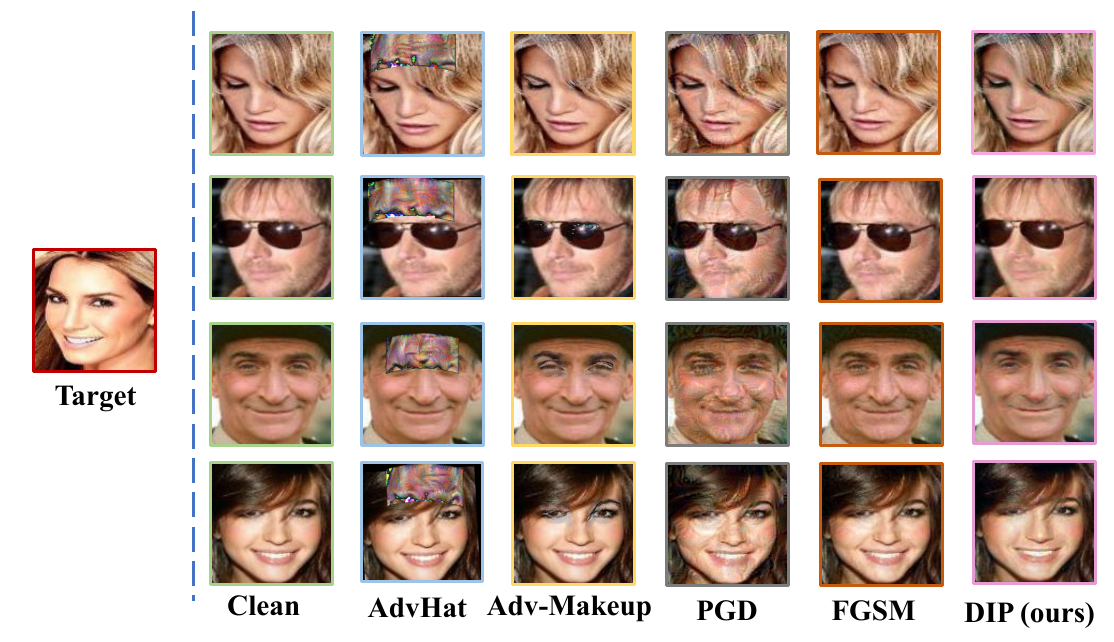}
   
        \caption{Visual comparison of adversarial examples and watermarked images obtained by different methods on the CelebA dataset.}	
 \label{img}
\end{figure}
\subsection{Robustness Analysis}
Online platforms typically process images to conform to standardized transmission and storage protocols. Therefore, the robustness of identity privacy protection schemes needs to be emphasized to ensure their feasibility in practical deployment. We applied Gaussian noise, JPEG compression (JPEG), resizing or Gaussian Blur to the generated images of all methods before evaluating their adversarial robustness and traceability robustness. As shown in Table~\ref{table2}, the ASR of adversarial attack methods significantly decreases after various image processing operations, with a maximum decrease of up to $40\%$ on the IRSE50 model after resizing, while the ASR of DIP-Watermark only decreases by approximately $20\%$, maintaining optimal performance across all evaluated victim models. Furthermore, after image processing, the traceability accuracy of DIP-Watermark consistently remains above 0.96, exceeding that of combined methods by up to $10\%$. This indicates the stable robustness of DIP-Watermark across multiple tasks. This robustness advantage can be attributed to two factors. Firstly, unlike adversarial attack methods based on iterative gradients and direct overlaying of perturbations on images, DIP-Watermark embeds watermark information into the deep feature space of the carrier, with the encoder learning to encode the carrier's general robust features. Additionally, DIP-Watermark optimizes the watermark encoder and decoder collaboratively while introducing adversarial transferability through meta-optimization strategies, thereby increasing the output dimensions of the model through multitask learning. ~\cite{mao2020multitask} has confirmed that multitask learning, especially joint optimization of low-correlation tasks, significantly enhances model robustness.

\subsection{Visualization and Analysis}

\begin{figure}
	\centering
	\includegraphics[width=86mm]{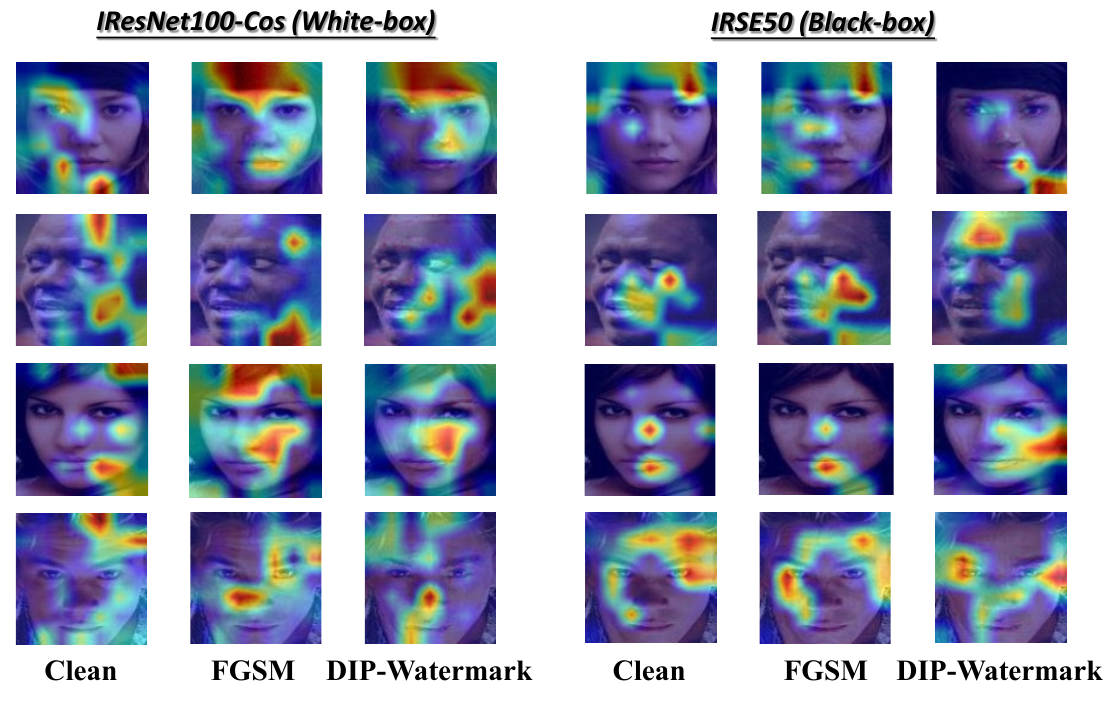}
   
        \caption{Visualization using Grad-CAM  produces attention maps on IResNet100-Cos model and IRSE50 model.}	
 \label{cam}
\end{figure}

We present the visual results of different methods using samples covering various genders, ages, and degrees of facial occlusion, as shown in Fig.~\ref{img}. Patch-based methods rely on significant semantic modifications to images. Adv-Makeup has limitations for images with occluded eyes. Perturbation-based methods produce adversarial examples with regular noise patterns, easy to be discovered and defended by adversaries.  DIP-Watermark-generated watermarked images exhibit only differences in overall hue, without noticeable artifacts. Additionally, we use Grad-CAM~\cite{selvaraju2017grad} visualization to illustrate the gradient responses of FGSM and  DIP-Watermark under black-box and white-box models. As shown in Fig.~\ref{cam}, FGSM and  DIP-Watermark both modify gradient responses in the white-box model. While FGSM overfit to modifications in the region of interest of the white-box model, including the background, resulting in poor transferability to black-box models.  DIP-Watermark encodes more universal adversarial features of the carrier, focusing on crucial facial regions, and effectively transfers these adversariality to the black-box model.




\begin{table}
	 
	\renewcommand\arraystretch{1.2}  
\resizebox{\hsize}{!}{
 
       \begin{tabular}{c|cc|cc|cc|cc|cc|cc}
       \Xhline{1px}
            \multirow{2}{*}{\textbf{Method}}&\multicolumn{2}{c}{\textbf{IR152}} &\multicolumn{2}{c}{\textbf{IRSE50}}&\multicolumn{2}{c}{\textbf{IResNet100-Arc}} &\multicolumn{2}{c}{\textbf{MobileFace}} & \multicolumn{2}{c}{\textbf{FaceNet}}  &\multicolumn{2}{c}{\textbf{IResNet100-Cos}} 
            \\ \cline{2-13}

			& ASR&SSIM& ASR&SSIM& ASR&SSIM& ASR&SSIM& ASR&SSIM& ASR&SSIM \\
            \hline
			w/o \textit{Meta} &0.136 &0.867 &0.522 &0.873 &0.193 &0.867 &0.601 &0.859 &0.496  &0.865  &0.227 &0.873 \\
			 DIP-Watermark  &\textbf{0.326} &\textbf{0.903} &\textbf{0.810} &\textbf{0.896} &\textbf{0.381} &\textbf{0.911}&\textbf{0.716} &\textbf{0.911} &\textbf{0.643}  &\textbf{0.898}  &\textbf{0.355} &\textbf{0.913} \\
			\hline
\Xhline{1px}
	\end{tabular}}
\caption{The effectiveness evaluation of meta-optimization methods. \textbf{w/o \textit{Meta}} indicates  DIP-Watermark is trained with a direct ensemble of FR models, without meta-optimization.}
	\label{t4}

\end{table}
\begin{table}
	 
	\renewcommand\arraystretch{1.2}  
\resizebox{\hsize}{!}{
 
       \begin{tabular}{c|cc|cc|cc|cc|cc|cc}
       \Xhline{1px}
            \multirow{2}{*}{\textbf{$\lambda _{adv}: \lambda _{en}:\lambda _{wm}$}}&\multicolumn{2}{c}{\textbf{IR152}} &\multicolumn{2}{c}{\textbf{IRSE50}}&\multicolumn{2}{c}{\textbf{IResNet100-Arc}} &\multicolumn{2}{c}{\textbf{MobileFace}} & \multicolumn{2}{c}{\textbf{FaceNet}}  &\multicolumn{2}{c}{\textbf{IResNet100-Cos}} 
            \\ \cline{2-13}

			& ASR&ACC& ASR&ACC& ASR&ACC& ASR&ACC& ASR&ACC& ASR&ACC \\
            \hline
			1:1:1 &0.025 &0.932 &0.125 &0.954 &0.005 &0.987 &0.006 &0.977 &0.000  &0.983  &0.000 &0.967 \\
   10:1:1  &0.100 &0.977 &0.092 &0.968 &0.000 &0.972 &0.167 &0.983 &0.063  &0.960  &0.000 &0.968 \\
			100:0.05:0.05  &\textbf{0.326} &\textbf{0.992} &\textbf{0.810} &\textbf{0.997} &\textbf{0.381} &\textbf{0.996}&\textbf{0.716} &\textbf{0.997} &\textbf{0.643}  &\textbf{0.994}  &\textbf{0.355} &\textbf{0.996} \\
			\hline
\Xhline{1px}
	\end{tabular}}
\caption{Impact of various weight coefficients on model performance.}
	\label{weight}

\end{table}

\subsection{Ablation Study}

To illustrate the influence of incorporating the meta-optimization strategy compatible with multiple tasks in end-to-end optimization on the comprehensive performance of  DIP-Watermark, we present the results of ablation experiments in Table~\ref{t4}. Meta-optimization strategy in  DIP-Watermark simultaneously enhances image quality and ASR. This is attributed to the encoder quickly adapting to new tasks with minimal gradient updates during meta-optimization, acquiring universal adversarial features for the carrier against unknown victim models. This avoids excessive reliance on penalizing image quality to enhance adversarial characteristics.

Additionally, we evaluate the impact of different weights on model performance, as shown in Table~\ref{weight}. As previously discussed, under equal weights, adversarial loss fails to converge after uniform training, resulting in poor black-box adversarial performance across all models. However, with the introduction of magnitude disparities, the training loss effectively converges, significantly enhancing adversarial performance, indicating the necessity of uniform loss value magnitudes during initial training stages.

\section{Conclusion}


To address the limitations of existing privacy protection methods, we propose  DIP-Watermark, the first method that protects facial image privacy from both adversarial and traceable perspectives. DIP-Watermark embeds a robust watermark into facial images, thwarting unauthorized malicious FR retrievals while enabling authorizers to extract the watermark for original identity tracing, thus mitigating potential threats of privacy leakage and identity impersonation. Specifically, we first propose an information-guided adversarial attack strategy, wich guides recognizable features of the image to deviate from the source identity features through an identity-specific watermark. Then we introduce a meta-optimization strategy that is compatible with multiple tasks to solve the optimization conflict between transferability and watermark invisibility. We conducted extensive experiments on two large-scale facial datasets, confirming that DIP-Watermark achieves significant adversarial transferability, traceability, and robustness against state-of-the-art face recognition models. It surpasses both adversarial attack-based methods and methods that simply combine adversarial attacks directly with watermarking.


%

\ifCLASSOPTIONcompsoc
  \section*{Acknowledgments}
\else
  \section*{Acknowledgment}
\fi
This work was supported by National Natural Science Foundation of
China NSFC (No. 62072343); the Fundamental Research Funds for the
Central Universities (No. 2042023kf0228); the National Key Research and
Development Program of China (No. 2019QY(Y)0206).

\bibliographystyle{IEEEtran}

\bibliography{ijcai23_0509}

\begin{thebibliography}{10}
\providecommand{\url}[1]{#1}
\csname url@samestyle\endcsname
\providecommand{\newblock}{\relax}
\providecommand{\bibinfo}[2]{#2}
\providecommand{\BIBentrySTDinterwordspacing}{\spaceskip=0pt\relax}
\providecommand{\BIBentryALTinterwordstretchfactor}{4}
\providecommand{\BIBentryALTinterwordspacing}{\spaceskip=\fontdimen2\font plus
\BIBentryALTinterwordstretchfactor\fontdimen3\font minus \fontdimen4\font\relax}
\providecommand{\BIBforeignlanguage}[2]{{%
\expandafter\ifx\csname l@#1\endcsname\relax
\typeout{** WARNING: IEEEtran.bst: No hyphenation pattern has been}%
\typeout{** loaded for the language `#1'. Using the pattern for}%
\typeout{** the default language instead.}%
\else
\language=\csname l@#1\endcsname
\fi
#2}}
\providecommand{\BIBdecl}{\relax}
\BIBdecl

\bibitem{liu2023diffprotect}
J.~Liu, C.~P. Lau, and R.~Chellappa, ``Diffprotect: Generate adversarial examples with diffusion models for facial privacy protection,'' \emph{arXiv preprint arXiv:2305.13625}, 2023.

\bibitem{hu2022protecting}
S.~Hu, X.~Liu, Y.~Zhang, M.~Li, L.~Y. Zhang, H.~Jin, and L.~Wu, ``Protecting facial privacy: Generating adversarial identity masks via style-robust makeup transfer,'' in \emph{Proceedings of the IEEE/CVF Conference on Computer Vision and Pattern Recognition}, 2022, pp. 15\,014--15\,023.

\bibitem{NEURIPS2023_29e4b51d}
H.~Otroshi~Shahreza and S.~Marcel, ``Face reconstruction from facial templates by learning latent space of a generator network,'' in \emph{Advances in Neural Information Processing Systems}, vol.~36, 2023, pp. 12\,703--12\,720.

\bibitem{9633176}
J.~Morris, S.~Newman, K.~Palaniappan, J.~Fan, and D.~Lin, ``“do you know you are tracked by photos that you didn’t take”: Large-scale location-aware multi-party image privacy protection,'' \emph{IEEE Transactions on Dependable and Secure Computing}, vol.~20, no.~1, pp. 301--312, 2023.

\bibitem{9442369}
J.~Lei, Q.~Pei, Y.~Wang, W.~Sun, and X.~Liu, ``Privface: Fast privacy-preserving face authentication with revocable and reusable biometric credentials,'' \emph{IEEE Transactions on Dependable and Secure Computing}, vol.~19, no.~5, pp. 3101--3112, 2022.

\bibitem{komkov2021advhat}
S.~Komkov and A.~Petiushko, ``Advhat: Real-world adversarial attack on arcface face id system,'' in \emph{2020 25th International Conference on Pattern Recognition}.\hskip 1em plus 0.5em minus 0.4em\relax IEEE, 2021, pp. 819--826.

\bibitem{jia2022adv}
S.~Jia, B.~Yin, T.~Yao, S.~Ding, C.~Shen, X.~Yang, and C.~Ma, ``Adv-attribute: Inconspicuous and transferable adversarial attack on face recognition,'' \emph{Advances in Neural Information Processing Systems}, vol.~35, pp. 34\,136--34\,147, 2022.

\bibitem{Li_2023_CVPR}
Z.~Li, B.~Yin, T.~Yao, J.~Guo, S.~Ding, S.~Chen, and C.~Liu, ``Sibling-attack: Rethinking transferable adversarial attacks against face recognition,'' in \emph{Proceedings of the IEEE/CVF Conference on Computer Vision and Pattern Recognition (CVPR)}, June 2023, pp. 24\,626--24\,637.

\bibitem{Dong2019EfficientDB}
Y.~Dong, H.~Su, B.~Wu, Z.~Li, W.~Liu, T.~Zhang, and J.~Zhu, ``Efficient decision-based black-box adversarial attacks on face recognition,'' \emph{2019 IEEE/CVF Conference on Computer Vision and Pattern Recognition}, pp. 7706--7714, 2019.

\bibitem{yang2021attacks}
L.~Yang, Q.~Song, and Y.~Wu, ``Attacks on state-of-the-art face recognition using attentional adversarial attack generative network,'' \emph{Multimedia tools and applications}, vol.~80, pp. 855--875, 2021.

\bibitem{wei2022simultaneously}
X.~Wei, Y.~Guo, J.~Yu, and B.~Zhang, ``Simultaneously optimizing perturbations and positions for black-box adversarial patch attacks,'' \emph{IEEE transactions on pattern analysis and machine intelligence}, 2022.

\bibitem{zhu2018hidden}
J.~Zhu, R.~Kaplan, J.~Johnson, and L.~Fei-Fei, ``Hidden: Hiding data with deep networks,'' in \emph{Proceedings of the European conference on computer vision}, 2018, pp. 657--672.

\bibitem{sharif2019general}
M.~Sharif, S.~Bhagavatula, L.~Bauer, and M.~K. Reiter, ``A general framework for adversarial examples with objectives,'' \emph{ACM Transactions on Privacy and Security (TOPS)}, vol.~22, no.~3, pp. 1--30, 2019.

\bibitem{10.1145/2976749.2978392}
{Sharif, Mahmood and Bhagavatula, Sruti and Bauer, Lujo and Reiter, Michael K.}, ``Accessorize to a crime: Real and stealthy attacks on state-of-the-art face recognition,'' in \emph{Proceedings of the 2016 ACM SIGSAC Conference on Computer and Communications Security}, 2016, p. 1528–1540.

\bibitem{8803269}
Z.~Zhu, Y.~Lu, and C.~Chiang, ``Generating adversarial examples by makeup attacks on face recognition,'' in \emph{2019 IEEE International Conference on Image Processing}, 2019, pp. 2516--2520.

\bibitem{yin2021adv}
B.~Yin, W.~Wang, T.~Yao, J.~Guo, Z.~Kong, S.~Ding, J.~Li, and C.~Liu, ``Adv-makeup: A new imperceptible and transferable attack on face recognition,'' \emph{arXiv preprint arXiv:2105.03162}, 2021.

\bibitem{madry2017towards}
A.~Madry, A.~Makelov, L.~Schmidt, D.~Tsipras, and A.~Vladu, ``Towards deep learning models resistant to adversarial attacks,'' \emph{arXiv preprint arXiv:1706.06083}, 2017.

\bibitem{goodfellow2014explaining}
I.~J. Goodfellow, J.~Shlens, and C.~Szegedy, ``Explaining and harnessing adversarial examples,'' \emph{arXiv preprint arXiv:1412.6572}, 2014.

\bibitem{zhong2020towards}
Y.~Zhong and W.~Deng, ``Towards transferable adversarial attack against deep face recognition,'' \emph{IEEE Transactions on Information Forensics and Security}, vol.~16, pp. 1452--1466, 2020.

\bibitem{wang2021faketagger}
R.~Wang, F.~Juefei-Xu, M.~Luo, Y.~Liu, and L.~Wang, ``Faketagger: Robust safeguards against deepfake dissemination via provenance tracking,'' in \emph{Proc. 29th ACM Int. Conf. Multimedia}, 2021, pp. 3546--3555.

\bibitem{liu2019novel}
Y.~Liu, M.~Guo, J.~Zhang, Y.~Zhu, and X.~Xie, ``A novel two-stage separable deep learning framework for practical blind watermarking,'' in \emph{Proceedings of the 27th ACM International conference on multimedia}, 2019, pp. 1509--1517.

\bibitem{9999475}
M.~He, H.~Wang, F.~Zhang, S.~M. Abdullahi, and L.~Yang, ``Robust blind video watermarking against geometric deformations and online video sharing platform processing,'' \emph{IEEE Transactions on Dependable and Secure Computing}, vol.~20, no.~6, pp. 4702--4718, 2023.

\bibitem{jia2021mbrs}
Z.~Jia, H.~Fang, and W.~Zhang, ``Mbrs: Enhancing robustness of dnn-based watermarking by mini-batch of real and simulated jpeg compression,'' in \emph{Proceedings of the 29th ACM international conference on multimedia}, 2021, pp. 41--49.

\bibitem{Yu_2021_ICCV}
N.~Yu, V.~Skripniuk, S.~Abdelnabi, and M.~Fritz, ``Artificial fingerprinting for generative models: Rooting deepfake attribution in training data,'' in \emph{Proceedings of the IEEE/CVF International Conference on Computer Vision}, 2021, pp. 14\,448--14\,457.

\bibitem{woo2018cbam}
S.~Woo, J.~Park, J.-Y. Lee, and I.~S. Kweon, ``Cbam: Convolutional block attention module,'' in \emph{Proc. Eur. Conf. Comput. Vis. (ECCV)}, 2018, pp. 3--19.

\bibitem{deng2019arcface}
J.~Deng, J.~Guo, N.~Xue, and S.~Zafeiriou, ``Arcface: Additive angular margin loss for deep face recognition,'' in \emph{Proceedings of the IEEE/CVF conference on computer vision and pattern recognition}, 2019, pp. 4690--4699.

\bibitem{mao2020multitask}
C.~Mao, A.~Gupta, V.~Nitin, B.~Ray, S.~Song, J.~Yang, and C.~Vondrick, ``Multitask learning strengthens adversarial robustness,'' in \emph{Computer Vision--ECCV 2020: 16th European Conference, Glasgow, UK, August 23--28, 2020, Proceedings, Part II 16}, 2020, pp. 158--174.

\bibitem{finn2017model}
C.~Finn, P.~Abbeel, and S.~Levine, ``Model-agnostic meta-learning for fast adaptation of deep networks,'' in \emph{International conference on machine learning}, 2017, pp. 1126--1135.

\bibitem{huang2008labeled}
G.~B. Huang, M.~Mattar, T.~Berg, and E.~Learned-Miller, ``Labeled faces in the wild: A database forstudying face recognition in unconstrained environments,'' in \emph{Workshop on faces in'Real-Life'Images: detection, alignment, and recognition}, 2008.

\bibitem{liu2015deep}
Z.~Liu, P.~Luo, X.~Wang, and X.~Tang, ``Deep learning face attributes in the wild,'' in \emph{Proceedings of the IEEE international conference on computer vision}, 2015, pp. 3730--3738.

\bibitem{7553523}
K.~Zhang, Z.~Zhang, Z.~Li, and Y.~Qiao, ``Joint face detection and alignment using multitask cascaded convolutional networks,'' \emph{IEEE Signal Processing Letters}, vol.~23, no.~10, pp. 1499--1503, 2016.

\bibitem{duta2021improved}
I.~C. Duta, L.~Liu, F.~Zhu, and L.~Shao, ``Improved residual networks for image and video recognition,'' in \emph{2020 25th International Conference on Pattern Recognition (ICPR)}.\hskip 1em plus 0.5em minus 0.4em\relax IEEE, 2021, pp. 9415--9422.

\bibitem{wang2018cosface}
H.~Wang, Y.~Wang, Z.~Zhou, X.~Ji, D.~Gong, J.~Zhou, Z.~Li, and W.~Liu, ``Cosface: Large margin cosine loss for deep face recognition,'' in \emph{Proceedings of the IEEE conference on computer vision and pattern recognition}, 2018, pp. 5265--5274.

\bibitem{MEGVII}
MEGVII, ``In https://www.faceplusplus.com.cn/,'' 2021.

\bibitem{Aliyun}
Aliyun, ``https://cn.aliyun.com/,'' 2019.

\bibitem{Wang_Bovik_Sheikh_Simoncelli_2004}
Z.~Wang, A.~Bovik, H.~Sheikh, and E.~Simoncelli, ``Image quality assessment: From error visibility to structural similarity,'' \emph{IEEE Trans. Image Process.}, p. 600–612, Apr 2004.

\bibitem{deb2020advfaces}
D.~Deb, J.~Zhang, and A.~K. Jain, ``Advfaces: Adversarial face synthesis,'' in \emph{2020 IEEE International Joint Conference on Biometrics}, 2020, pp. 1--10.

\bibitem{selvaraju2017grad}
R.~R. Selvaraju, M.~Cogswell, A.~Das, R.~Vedantam, D.~Parikh, and D.~Batra, ``Grad-cam: Visual explanations from deep networks via gradient-based localization,'' in \emph{Proceedings of the IEEE international conference on computer vision}, 2017, pp. 618--626.

\end{thebibliography}

\begin{IEEEbiography}
[{\includegraphics[width=1in,height=1.25in,clip,keepaspectratio]{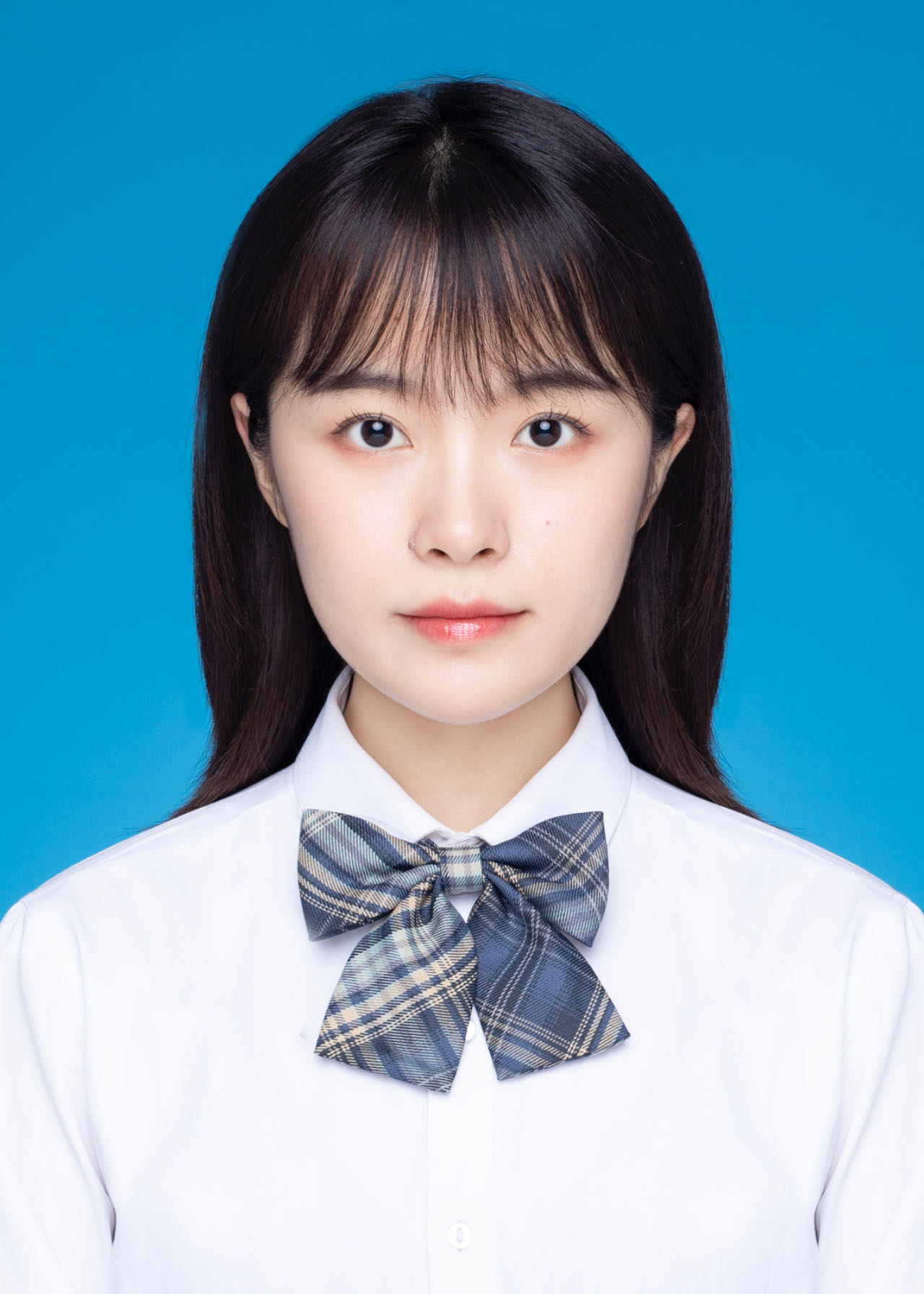}}] 
{Yunming Zhang} received the MS degree from the School of Information Science and Engineering, Shandong Normal University, Jinan, China, in 2022. She is currently working toward the doctor’s degree in the School of Cyber Science and Engineering, Wuhan University, Wuhan, China. She research interests include Artificial intelligence security, multimedia forensics, and deepwatermarking.
\end{IEEEbiography}

\begin{IEEEbiography}[{\includegraphics[width=1in,height=1.25in,clip,keepaspectratio]{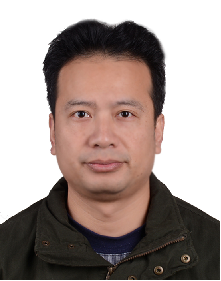}}]{Dengpan Ye}(Member, IEEE) received the B.Sc. degree in automatic control from SCUT in 1996 and the Ph.D. degree from NJUST in 2005. He was a Post-Doctoral Fellow in information system with Singapore Management University. Since 2012, he has been a Professor with the School of Cyber Science and Engineering, Wuhan University. He has authored or coauthored over 30 refereed journal and conference papers. His research interests include machine learning and multimedia security.
\end{IEEEbiography}

\begin{IEEEbiography}
[{\includegraphics[width=1in,height=1.25in,clip,keepaspectratio]{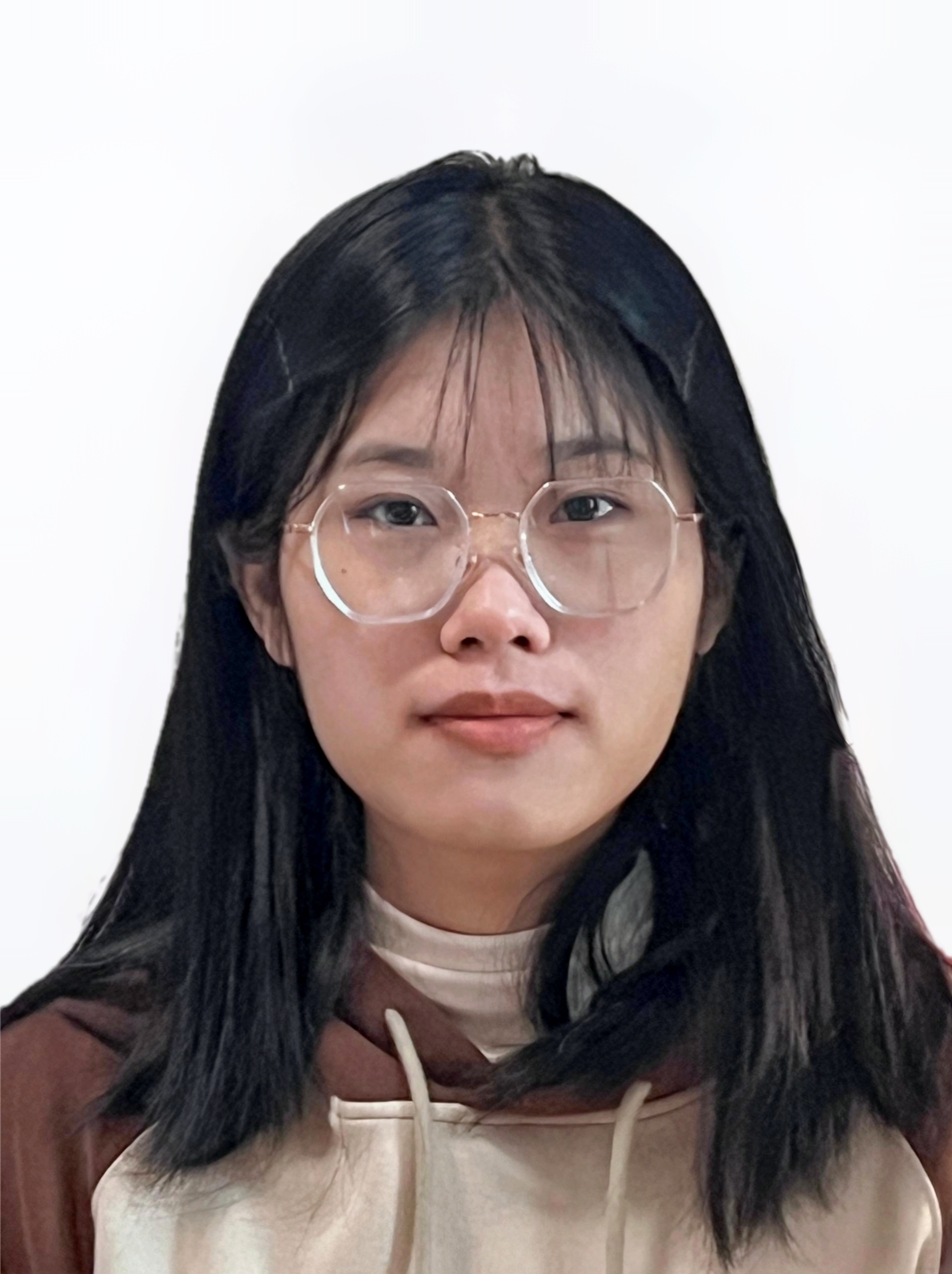}}] 
{Xie Caiyun} received her bachelor's degree from the School of Computer Science, Wuhan University, Hubei, China, in 2023. She is currently pursuing a master's degree in the School of Cyber Science and Engineering, Wuhan University, Wuhan, China. Her research interests include artificial intelligence security and deep watermarking technology.
\end{IEEEbiography}

\begin{IEEEbiography}
[{\includegraphics[width=1in,height=1.25in,clip,keepaspectratio]{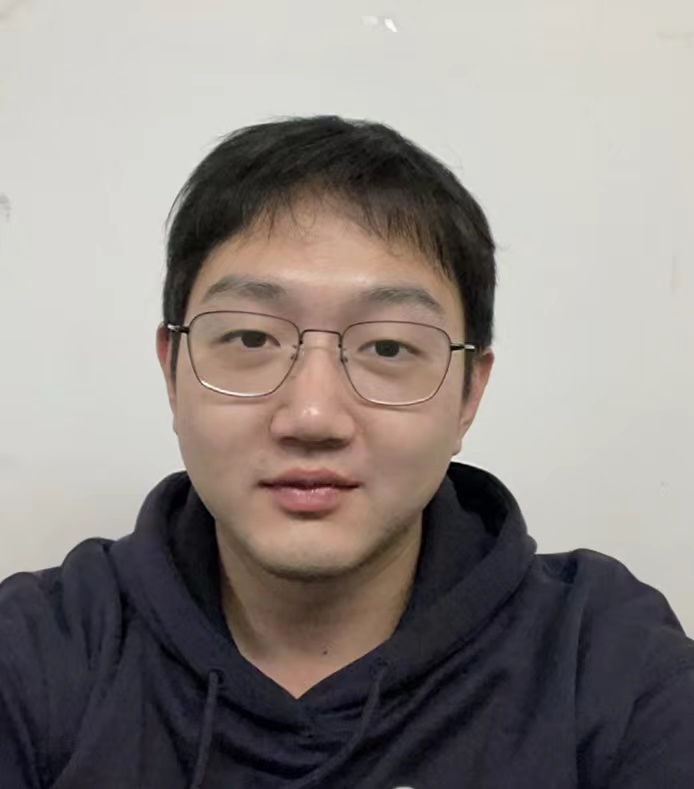}}] 
{Sipeng Shen} received the B.S. degree of Information Security in Software College, Northeastern University, Liaoning, China. He is currently pursuing a master's degree in the School of Cyber Science and Engineering, Wuhan University, Wuhan, China. His research interests include Information Security, Multimedia Security and Deep Learning.
\end{IEEEbiography}

\begin{IEEEbiography}
[{\includegraphics[width=1in,height=1.25in,clip,keepaspectratio]{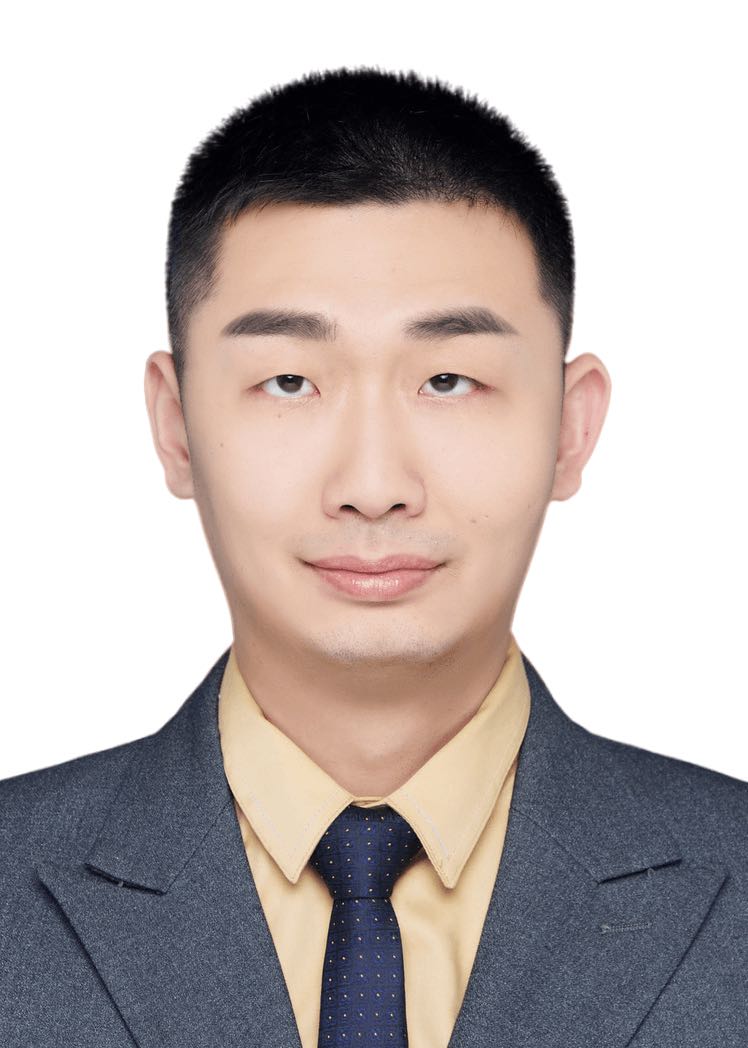}}] 
{Ziyi Liu} received the B.S. degree from Chongqing Jiaotong University, Chongqing, China, in 2017, and the M.S. degree from the Guilin University Of Electronic Technology, Guilin, China, in 2022. Currently studying for a PhD at Wuhan University. His research interests include machine learning, network security, and federated learning.
\end{IEEEbiography}

\begin{IEEEbiography}
[{\includegraphics[width=1in,height=1.25in,clip,keepaspectratio]{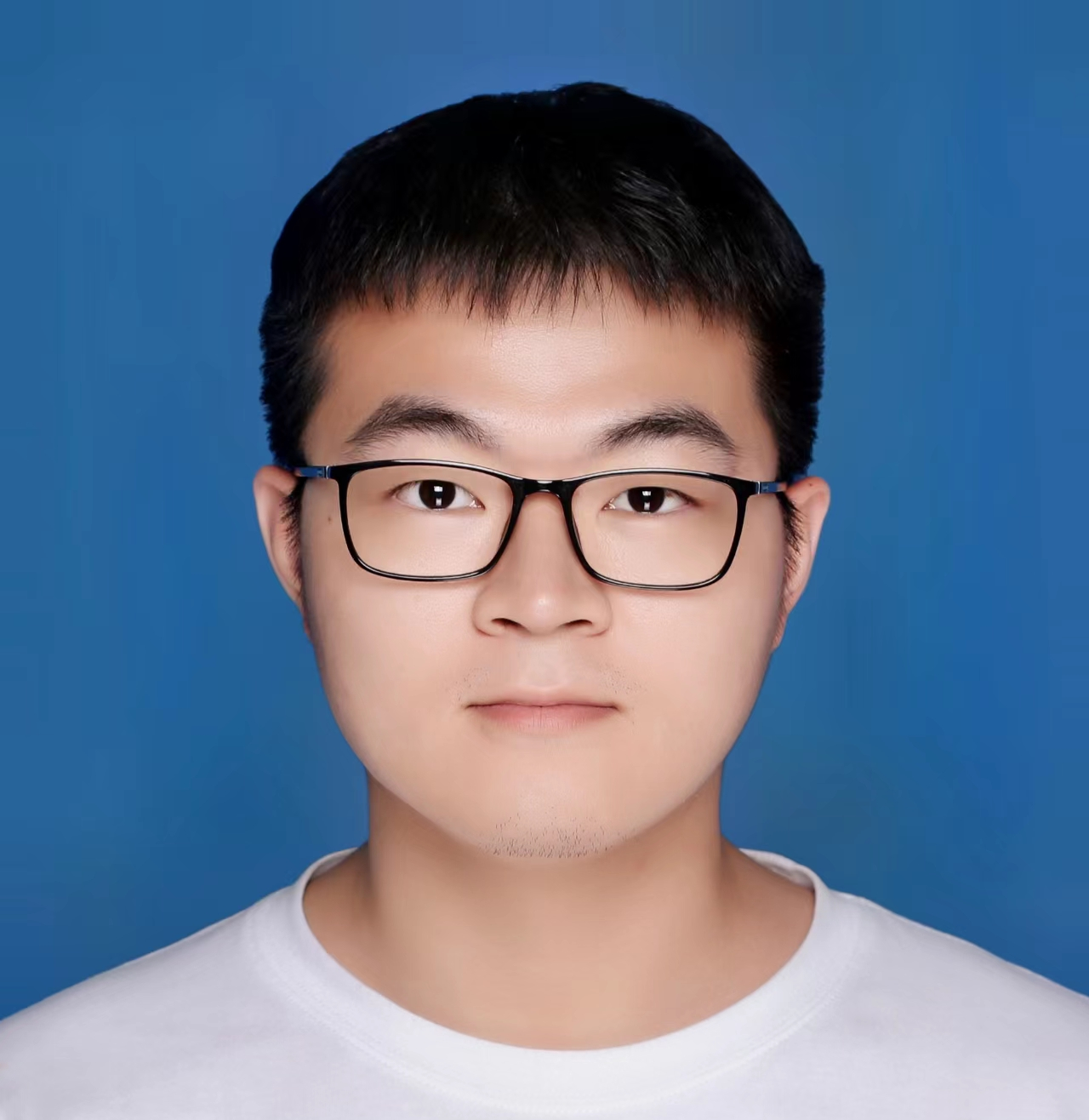}}] 
{Jiacheng Deng} received the MS degree in computer science from Ningbo University, Ningbo, P.R. China, in 2023. He is currently working toward the doctor’s degree in the School of Cyber Science and Engineering, Wuhan University, Wuhan, P.R. China. His research interests include theories in Synthetic Speech Detection, Adversarial attacking, and Explanation of Neural Networks.
\end{IEEEbiography}

\begin{IEEEbiography}
[{\includegraphics[width=1in,height=1.25in,clip,keepaspectratio]{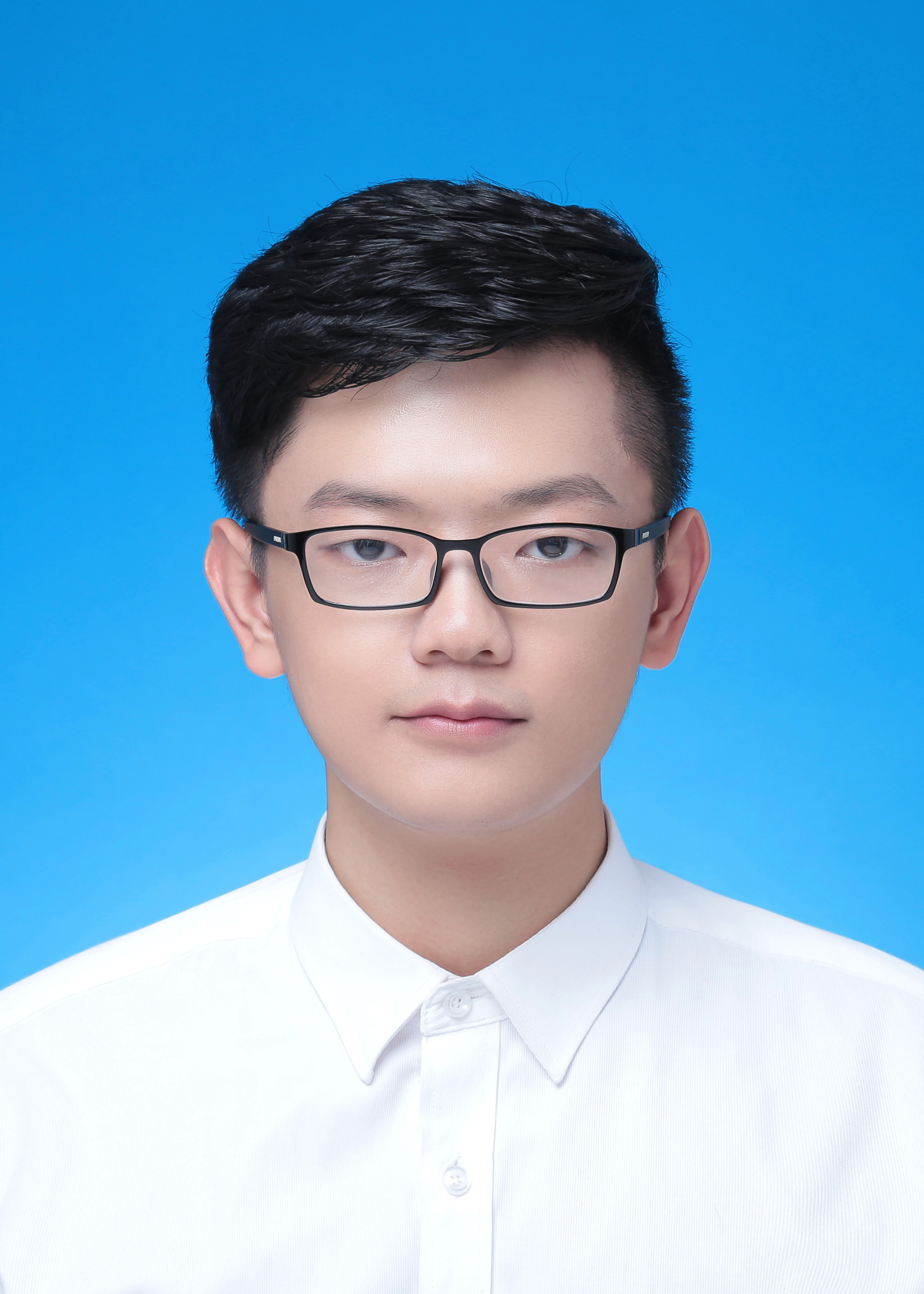}}] 
{Long Tang} received the MS degree in computer science from Xidian University, Xi’an, P.R. China, in 2021. He is currently working toward the doctor’s degree in the School of Cyber Science and Engineering, Wuhan University, Wuhan, P.R. China. His research interests include theories in Adversarial attacking and defending, DeepFake generation and detection, and Verification of neural networks.
\end{IEEEbiography}

%







\end{document}